\documentclass[usenatbib]{mnras}
\usepackage[utf8]{inputenc}
\usepackage{amsmath}
\usepackage{graphicx}
\usepackage[T1]{fontenc}
\usepackage{ae,aecompl}
\usepackage{subfig}
\usepackage{rotating}

\title[Illustris disc galaxies formed from major mergers]{Disc galaxies formed from major mergers in Illustris}

\author[N. Peschken, E. L. {\L}okas and E. Athanassoula]{Nicolas Peschken$^{1}$\thanks{Contact e-mail:
\href{mailto:npeschken@camk.edu.pl}{npeschken@camk.edu.pl}}, Ewa L. {\L}okas$^{2}$ and E. Athanassoula$^{3}$\\
$^{1}$Centre for Astronomy, Nicolaus Copernicus University, Faculty of Physics, Astronomy and Informatics,
Grudziadzka 5, 87-100 Toru\'{n}, Poland \\
$^{2}$Nicolaus Copernicus Astronomical Center, Polish Academy of Sciences, Bartycka 18, 00-716 Warsaw, Poland \\
$^{3}$Aix Marseille Univ, CNRS, CNES, LAM, UMR 7326, F-13388 Marseille 13, France
}

\pubyear{2018}

\begin{document}
\label{firstpage}
\pagerange{\pageref{firstpage}--\pageref{lastpage}}
\maketitle

\begin{abstract}
 We show how wet major mergers can create disc galaxies in a cosmological context, using the Illustris simulation. We
select a sample of 38 disc galaxies having experienced a major merger in their history with no subsequent significant minor merger,
and study how they transform into discs after the merger. In agreement with what was previously found in controlled simulations of such mergers, we find that their disc is built gradually
from young stars formed after the merger in the disc region, while the old stars born before the merger form an
ellipsoidal component. Focusing on one fiducial case from our sample, we show how the gas was initially dispersed in the
halo region right after the merger, but is then accreted onto a disc to form stars, and builds the disc component. We then
select a sample of major mergers creating elliptical galaxies, to show that those cases correspond mainly to dry
mergers, where the lack of star formation prevents the formation of a disc in the remnant galaxy. The amount of gas in
the remnant galaxy after the merger is therefore essential to determine the final outcome of a major merger.
\end{abstract}


\section{Introduction}

The major merger of two galaxies has long been thought to produce elliptical galaxies (since
\citealt{1972ApJ...178..623T}), and is even considered as the main mechanism responsible for the creation of
ellipticals. Colliding two similar mass spiral galaxies will often completely destroy their disc stellar structures,
scatter their stars and eventually assemble them in a spheroidal or ellipsoidal component
(\citealt{1977egsp.conf..401T, 1992ARA&A..30..705B, 2009A&A...496..381H, 2013ApJ...778...61T,2017MNRAS.467.3934D}) characteristic of
elliptical galaxies. This picture seemed very convincing, until in the last two decades both observations and
simulations showed that disc galaxies could also be created from a major merger. \cite{2005A&A...430..115H,
2009A&A...496..381H} found late type galaxies showing traces of a past major merger, while simulations (e.g.
\citealt{2005ApJ...622L...9S, 2006ApJ...645..986R, 2007MNRAS.374.1479G, 2009ApJ...691.1168H, 2014A&A...570A.103B}) showed that the remnant
of a major merger could have a small disc component, particularly if the two initial discs were on, or at least close to, the orbital plane.

\cite{2016ApJ...821...90A} (see also \citealt{2017A&A...600A..25R, 2017MNRAS.468..994P, 2018IAUS..334...65A}) showed
how to create a spiral galaxy from a major merger, using simulations of an isolated pair of disc galaxies with a hot
gaseous halo each, colliding at intermediate or high redshift. While the stars pile up in the center of the remnant to form a classical
bulge, gas is accreted from the halo after the merger, and allows the formation of a dominating disc component. This
paper was a proof of concept that major mergers can produce disc galaxies in the idealized scenario of two isolated
galaxies, and allowed to study and follow in detail and in a controlled environment the formation of the disc. Similar
work then followed, such as \cite{2017A&A...604A.105T}, \cite{2018A&A...617A.113E} and \cite{2018MNRAS.473.2521S}, who
confirmed these results. However, such simulations are run in isolation, therefore they neglect the effect of the
environment, and make assumptions on the parameters of the merger such as the orbit, or on the structures of the two
merging galaxies. It is therefore necessary to check whether the scenario of two merging galaxies leading to a disc
galaxy also works in a more general and realistic case of cosmological context, where the environment is expected
to play a strong role in galaxy evolution, and less assumptions are made on galaxies and their interactions.

An efficient way to do so is to use cosmological simulations, which have been intensively developed over the last
decade, and allow to study the formation and evolution of galaxies in a significant portion of the
Universe. Illustris-1 (hereafter Illustris) is a large high resolution hydrodynamical simulation
(\citealt{2014MNRAS.444.1518V}) modeling a cube of (106 Mpc)$^3$, enough to contain hundreds of thousands of galaxies,
and therefore constitutes a great laboratory for the study of galaxy evolution and interactions. It is based on the
AREPO code (\citealt{2010MNRAS.401..791S}), and has been shown to reproduce many observational results in the field of
galaxy evolution (e.g. \citealt{2015MNRAS.454.1886S}). Illustris contains about 18 billions of particles modelling dark
matter, gas and stars, with a mass resolution of 6.3 $\times$ 10$^6$ $M_{\odot}$ for the dark matter and 1.6 $\times$
10$^6$ $M_{\odot}$ for the gas, while the linear resolution is 0.7 kpc for baryonic matter, and 1.4 kpc for dark matter. The simulation is publicly available through 135 snapshots from redshift 127 to
the present time. As the number of particles in Illustris galaxies is about an order of magnitude lower than in the previously mentioned simulations run in isolation, we will look here only at global properties of the disc (i.e. its existence, or not), and do not extend our study to the disc properties and substructures.

A first attempt to work on the formation of disc galaxies from major mergers in Illustris has been undertaken by
\cite{2017MNRAS.470.3946S}, who studied four galaxies formed from major mergers with zoom-in simulations. They found
that, as already shown in controlled simulations, a disc component could be regrown in galaxies in a cosmological framework, after a major merger under certain feedback conditions in those
zoom-in simulations, and that star formation could still occur after the merger. However they focused mainly on the
circumstances of the quenching of the remnant galaxy, in particular on the effect of AGN feedback. We would like to
extend that work for a larger sample of Illustris galaxies fully evolved in a cosmological context, and track the
formation of the disc more closely, in particular including the role of the gas and its amount. This will allow us to
understand better in which scenarios major mergers produce ellipticals or spiral galaxies in a cosmological context
using statistics of a larger number of galaxies.

The paper is organized as follows. In section \ref{disc_glob} we select a sample of disc galaxies formed from a major
merger, and study their formation and evolution after the merger. Then in section \ref{fiduc} we focus on one
particular case from our sample, our fiducial example, and study in more detail how the disc is formed and the role of
the gas. In section \ref{ell_glob} we construct a sample of elliptical galaxies formed from a major merger, to compare
to our disc sample. We discuss the results in section \ref{discuss}, and conclude in section \ref{ccl}.

\section{Forming Disc galaxies from major mergers}
\label{disc_glob}

\subsection{Disc galaxies sample}
\label{disc_samp}

We are interested in major mergers producing disc galaxies in Illustris. To find such cases, we select at redshift
$z=0$ all disc galaxies and trace back their history to look for major mergers. To do this, we use a similar approach
as in \cite{2019MNRAS.483.2721P}, where we used the criteria of the flatness of the galaxy, as well as the circularity
of the stellar particles, to define disc galaxies. The circularity parameter $\epsilon$ of a stellar particle
represents how close its orbit is to a circular orbit, and is defined for a given particle as the ratio between the $z$-component of its angular momentum ($z$ being the rotation axis of the galactic disc), and the angular momentum of the circular orbit of the same radius. Disc particles are therefore expected to have $\epsilon$
close to 1. Illustris provides for every galaxy the fractional mass $f_{\epsilon}$ of stars with $\epsilon > 0.7$, and
we take $f_{\epsilon} > 0.2$ to define disc galaxies. We also use the flatness of galaxies as provided by Illustris
from the mass tensor, and set the additional constraint of $flatness < 0.7$ for a galaxy to be considered as a disc.

Furthermore, we keep only galaxies with more than 60 000 stellar particles, which corresponds to $M_*=5 \times 10^{10}
M_{\odot}$. This ensures that the stellar population is big enough to be able to probe its age and distribution
properly, which will be necessary to track the formation of a disc component.

\subsection{Major mergers sample}
\label{disc_sample_merg}

\begin{figure}
  \center
  \includegraphics[scale=0.5]{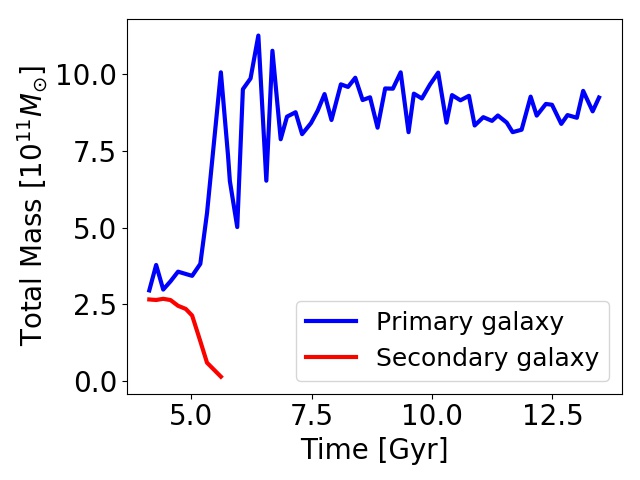}\\
\hspace{0.25cm}
  \includegraphics[scale=0.482]{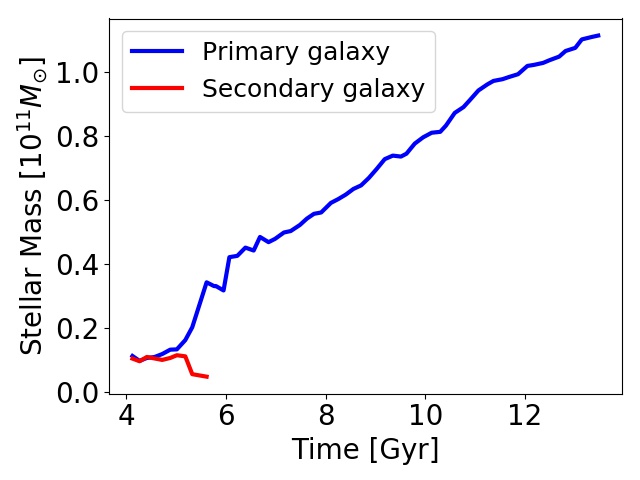}
  \caption{Total masses of both merging galaxies (top), as well as stellar masses (bottom panel), as a function of
time. Here the time represents the age of the Universe in Illustris.
  }
  \label{mass_16948}
\end{figure}

We now trace back the merger history of those selected disc galaxies starting from redshift $z=0$ up to redshift $z=1.5$, to
look for major mergers. To do this, we use the \textit{Sublink} merger tree of Illustris, which allows to look at each
snapshot for the progenitors of a given galaxy in the previous snapshot. Therefore, having the number of progenitors
higher than one at a given snapshot means that a merger occurred, and we select the cases with two progenitors to focus on the mergers of two galaxies. We are looking for major mergers, and keep therefore
only mergers with a total mass ratio higher than 0.25. However the masses provided by Illustris cannot be fully trusted,
since when two galaxies are close, Illustris struggles to attribute particles correctly to each galaxy. Taking
Illustris masses at the moment of the merger would often result in having one galaxy taking most of the mass, and the
other one having very little mass left, making it very difficult to find a major merger (see Fig. \ref{mass_16948} for
an example). It is thus needed to compute the masses of the galaxies before the two merging galaxies get too close to
know what is the mass ratio between the two progenitors. We dismiss the snapshots corresponding to the merger period,
where the galaxies are too close, and keep only snapshots before the merger where the distance between the two
progenitors is higher than 80 kpc. Even so, the total mass can sometimes vary due to the presence of other galaxies, so we choose to average the mass of each progenitor over 3 snapshots, to have a more reliable estimate of the real mass. We find that in most cases the mass values within those 3 snapshots do not differ by more than 50\% difference from the average value. This average mass then allows us to derive the mass ratio. 

To determine the moment of the merger, we cannot rely on the Illustris merger tree, as when two galaxies are too close
Illustris tends to consider they are one single galaxy, while the merger might still be ongoing. We therefore probed by
eye the snapshot at which the merger seems to start (i.e. when the two galaxies collide for the first time), as well as
the snapshot at which the merger seems to be over (i.e. when there is no more overdensity moving around the center of
mass, and the remnant seems to have stabilized). This gives us the merger period.

We now have for every disc galaxy of our sample the snapshots where a major merger (mass ratio > 0.25) occurred in its
history. However after a major merger, other events such as minor or other major mergers can occur, which can impact
the formation of a disc. We therefore keep only cases where after the major merger no significant external event occurred for the
remnant galaxy. To do this, we discard all the cases where a minor merger with a mass ratio higher than 5\% occurs
after the major merger. We also dismiss the cases where the major merger is a multiple merger, with other galaxies
involved. In case there are several major mergers in the history of the galaxy, we keep only the latest one.
This allows us to isolate the effect of the major merger on the subsequent evolution of the galaxy.

\begin{figure}
  \includegraphics[scale=0.5]{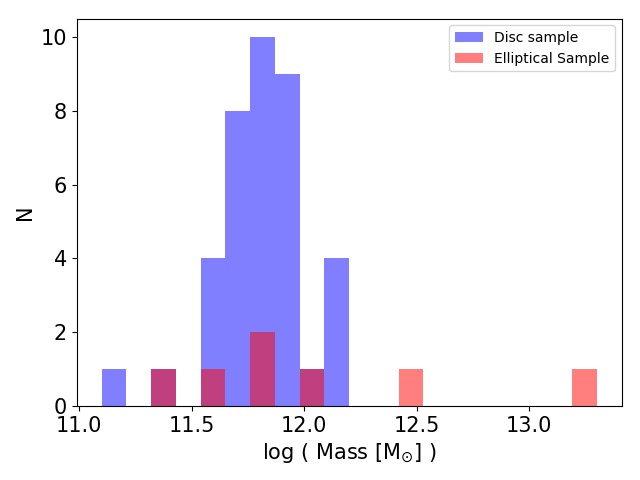}
  \includegraphics[scale=0.5]{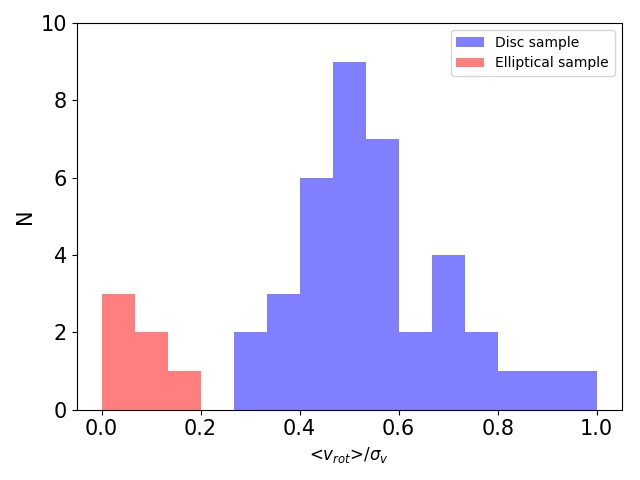}
  \includegraphics[scale=0.5]{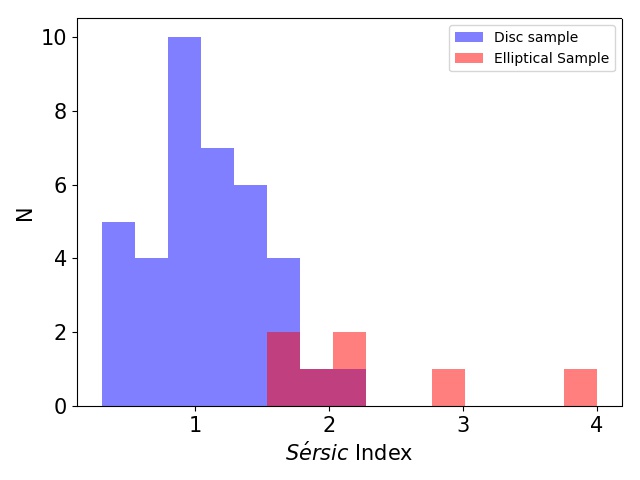}
\caption{Top panel: total masses of the galaxies in the disc sample (section \ref{disc_glob}) and in the elliptical sample
(section \ref{ell_glob}). Lower panels: histograms for both samples of quantities related to the discy nature of the galaxies at redshift $z=0$: rotation to dispersion ratios (middle panel), and S\'ersic indexes derived from the fit of the surface density profiles (bottom panel). These two panels show clearly that the disc sample is composed of galaxies that are more rotationally supported, and with a density profile closer to an exponential (S\'ersic index $\sim$ 1), than the galaxies of the elliptical sample.}
  \label{histg}
\end{figure}

In this way we obtain 38 disc galaxies with more than 60 000 stellar particles at redshift $z=0$, and have a clear major
merger in their history with no significant following merger. We call this sample our disc sample. The list of all galaxies in our sample together with their properties is given in the Appendix (Table \ref{table1}). As shown in
\cite{2019MNRAS.483.2721P}, the mass provided by Illustris cannot always be trusted, so we compute the total mass of
the galaxies in our sample ourselves, in the same way as in \cite{2019MNRAS.483.2721P}. We fit an exponential disc to
the stellar distribution of those galaxies within two stellar half-mass radii, and then take the sum of the
total mass inside 20 times the derived exponential scale length to derive the total mass, and 10 times to derive the
stellar mass.
We obtain an average total mass of $(7.24 \pm 3.10) \times 10^{11} M_{\odot}$ for our sample of galaxies at
redshift $z=0$, with their mass distribution displayed in Fig. \ref{histg} (top panel), and a corresponding stellar mass of
$(7.95 \pm 3.57) \times 10^{10} M_{\odot}$.

To verify the reliability of our disc sample, we checked if the $z=0$ galaxies of our sample are indeed realistic disc galaxies, i.e. if the criteria we have chosen to define a disc galaxy (circularity and flatness, see section \ref{disc_samp}) are accurate. We use two more observables to check how discy our galaxies are: their rotation and their surface density profile. We expect our disc galaxies to have a high rotation to velocity dispersion ratio, contrary to early type galaxies which are usually dispersion dominated. We therefore computed the mean tangential velocity in the disc inside twice the half stellar mass radius, and divided it by the velocity dispersion inside the same radius. We find for our disc sample  values around 0.6 (see Fig. \ref{histg}, middle panel), way higher than our elliptical sample (described in section \ref{ell_glob}), which confirms that our galaxies are rotationnally supported, in agreement with what we expect for disc galaxies. The values of the rotation to dispersion for all galaxies in our sample can be found in the Appendix (Table \ref{table1}).
  
  Furthermore, disc galaxies are expected to have a surface density profile resembling an exponential (\citealt{1970ApJ...160..811F}). To see whether this is the case in our sample, we computed those profiles for each galaxy at $z=0$ and fitted them with a S\'ersic profile: $\Sigma_{0}e^{-(R/R_S)^{1/n}}$, $R_S$ being the scale length. The S\'ersic index $n$ here tells us if the galaxy profile is close to an exponential disc ($n=1$) or to a de Vaucouleurs profile ($n=4$, \citealt{1948AnAp...11..247D}) describing rather bulge dominated galaxies (e.g. \citealt{2007ApJ...664..640D}). We performed those fits for each galaxy and find values around 1 (Fig. \ref{histg}, bottom panel, values in the Appendix, table \ref{table1}), distinct from elliptical galaxies and in good agreement with expectations for discs. Therefore, the galaxies in our sample at $z=0$ seem indeed to be reasonable disc galaxies from their surface density profile and their rotation, and distinct from our elliptical sample (see section \ref{ell_samp} for details).

\begin{figure}
  \includegraphics[scale=0.5]{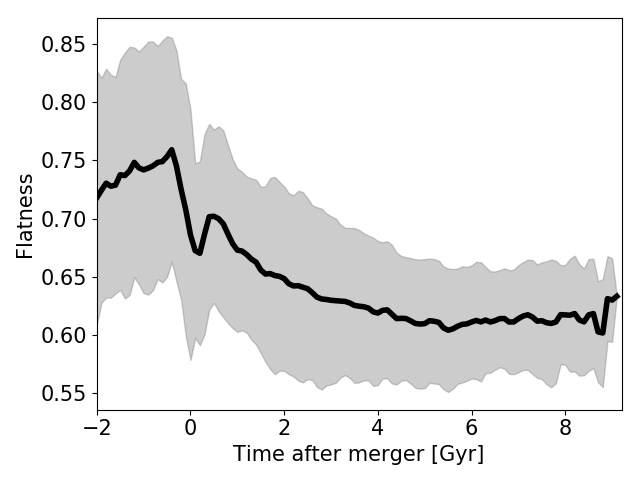}
\caption{Mean flatness (as defined in section \ref{res_disc}) as a function of time in our disc sample,
with the 1-$\sigma$ dispersion in grey. The flatness decreases over time, which means that the galaxy becomes flatter and
more disc-like. Since the merger occurs at different times for each galaxy of the sample, we scaled the time with
respect to the merger time, and used an interpolation scheme as described in section \ref{res_disc} to convert snapshots to time for the whole sample. The dispersion goes to zero at the final snapshot because there is only
one galaxy in our sample that has a merger as early as 60 snapshots before the end of the simulation.}
  \label{flat_samp}
\end{figure}

\subsection{Results}
\label{res_disc}

This sample of 38 galaxies is thus composed of disc galaxies having experienced a major merger, and thus shows how a
dominant disc can be built after a major merger, instead of an elliptical galaxy. This confirms the previous work on
the topic (e.g. \citealt{2016ApJ...821...90A}) where controlled simulations were used. Here we will investigate how and
why this disc is built in cosmological simulations.

In all cases, if there was a disc before the major merger, it was destroyed by the merger, there is no survival of the
disc. Therefore the disc can only have been formed after the merger. We will thus follow the evolution and
properties of the galaxies in our sample after the merger.

\begin{figure}
  \includegraphics[scale=0.5]{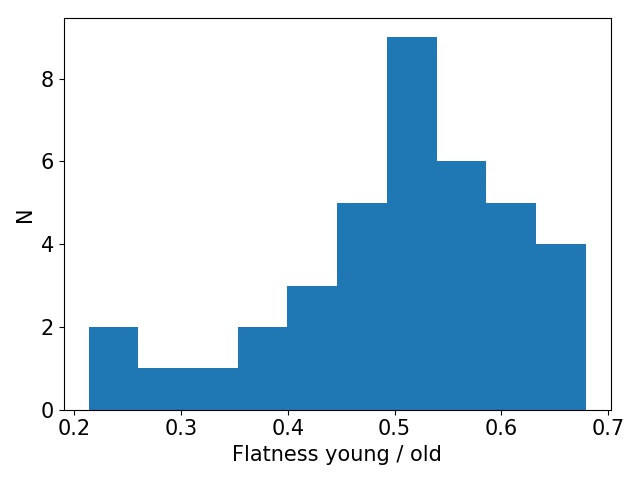}
\caption{Ratio of the flatness (as defined in section \ref{res_disc}) of the young to the old stellar population in
our disc sample, the old population being composed of the stars born before the merger, and the young one of those
born after. We see that the young population appears to be flatter than the old one.}
  \label{flat_ratio}
\end{figure}

To characterize the formation of the disc, we first look at the evolution of its shape, using the flatness of the
galaxy. Illustris provides the eigenvectors of the mass distribution, which allows us to derive the flatness of the
stellar component. However, as stated in the previous section, Illustris does not always attribute particles to galaxies
correctly (see also the Appendix in \citealt{2019MNRAS.483.2721P}), so these eigenvectors might be biased. We
therefore derive the flatness ourselves, using only the stellar particles inside twice the half-mass radius, which
we also computed ourselves. We compute the mass tensor $M$ of the stellar distribution ourselves, using:

\begin{equation}
  M(i,j)=\sum_p m_p X_{i,p} X_{j,p}
\end{equation}

\noindent with ${i,j}$ between 1 and 3 representing the 3 spatial directions x,y,z, and $X_{i,p}$ being the coordinate of the particle $p$ in the $i$ direction. We then define the flatness as $M_Z/\sqrt{M_X \times M_Y}$, where $M_X$, $M_Y$ and $M_Z$ are the
eigenvalues of the mass tensor. The lower the flatness, the flatter the galaxy is.

Following the flatness over time for all galaxies in our sample, we find a decrease of the flatness over time,
after the merger. We plot in Fig. \ref{flat_samp} the mean and dispersion of the flatness as a function of time, for
our whole sample. We start the plot 10 snapshots before the beginning of the merger to have a time consistent for all
cases, as each merger of the sample occurs at a different time. Since the time step between two snapshots is not
constant in Illustris, we cannot directly convert the number of snapshots to a time duration, and to obtain the flatness as a function of time averaged for all the galaxies in our sample, we had to perform an interpolation for each profile.

  To do this, for every galaxy of the sample we converted the snapshot numbers to time (i.e. age of the Universe), rescaled the time profile with respect to the merger time, and then used a spline function of order 3 for the interpolation. We obtain this way a continuous function of time for every galaxy, which is then averaged over the whole sample in Fig. \ref{flat_samp}, and in the following plots displaying time evolution. Although this interpolation smoothens slightly the profiles, it constitutes a good approximation of the time evolution of quantities rescaled with respect to the merger time, and allows us to have a consistent average of our sample as a function of time, instead of snapshot numbers.

Before the merger, the flatness is rather high, indeed the main progenitors are then often not disc galaxies, but rather small clumpy ellipsoidal shaped galaxies. We see that at the moment of the merger, the flatness increases due to the destruction of the galaxy, but then abruptly
drops, and later decreases gradually over time. It thus seems that the disc starts forming quickly after the merger, although the Illustris results do not allow us to give a precise quantitative estimate of the time-scale. In Fig. \ref{flat_samp} the time-scale is not consistent for all cases of the sample, with relative differences to the mean time we took (top axis) going up to 30\% between two time-steps, therefore some small shifts on the X-axis from one case to the other could slightly affect the appearance of the mean curve (for example the sharpness of the drop might be slightly different from one case to the other). However the results can be trusted qualitatively, and allow us to understand the general behaviour of the galaxies in our sample.
Note that before the merger occurs, we focus on the primary galaxy, i.e. the main progenitor as determined by the
Illustris merger tree (often the galaxy with a higher mass).

\begin{figure}
  \includegraphics[scale=0.5]{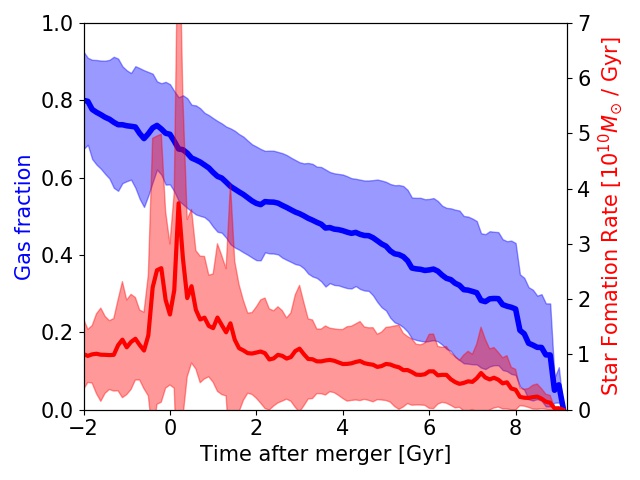}
\caption{Star formation rate and gas fraction as a function of time in our disc sample. As in Fig. \ref{flat_samp}, the
time is scaled with respect to the time of the merger for each galaxy.}
  \label{sfr_samp}
\end{figure}

To see whether the decrease in flatness comes from a rearrangement of the old stars, or the formation of a young disc,
we will now divide the stars at redshift $z=0$ into two groups. We define the old stars as all the stellar particles
born before the beginning of the merger, and the young stars as the ones born after the end of the merger. We exclude
the stars born during the merger, as the information given by the stars born during this violent process might not be easy to interpret in terms of disc formation. We find that the old stars form a rather spheroidal component, while the
younger stars form a clear disc component (see section \ref{fiduc} for a fiducial example).

\begin{figure}
  \includegraphics[scale=0.5]{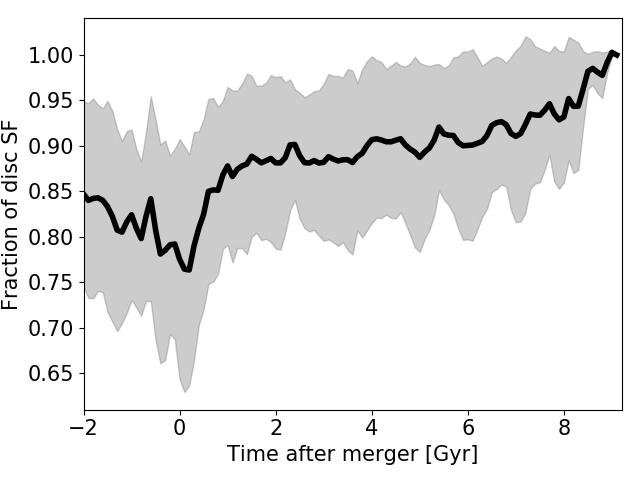}
\caption{Fraction of star formation occurring in the disc as a function of time in our sample of disc galaxies, i.e.
the mass of stars formed in the disc ($|Z|<5$ kpc, $R<50$ kpc), divided by the total mass of the stars formed in the
galaxy, at each time-step. Here again the snapshots are scaled with respect to the moment of the merger, as in Fig.
\ref{flat_samp} and \ref{sfr_samp}.}
  \label{SFdisc_samp}
\end{figure}

To characterize this, we compute the flatness (as defined above) separately for the young and old components, and look
at the ratio of its values for the young and the old population. We plot the histogram of the ratio distribution in
Fig. \ref{flat_ratio}, and find a mean value of 50 $\pm$ 12 \%, which shows that the young population is indeed
clearly flatter than the old one (in agreement with the results of \citealt{2016ApJ...821...90A}). Furthermore, while
the young population has a flatness of 0.38 $\pm$ 0.10, the old population shows a mean value of 0.76 $\pm$ 0.076, i.e.
close to the value the primary galaxies had before the merger (see Fig. \ref{flat_samp}). Therefore, the origin of
the disc is related to the young population formed after the merger, rather than the flattening of the old one.

As young stars seem central to understand the disc formation, the next logical step is to look at
the star formation rate over time in our galaxies. In Fig. \ref{sfr_samp} we plot over time the mean and the dispersion
values of the stellar mass formed in between two snapshots inside twice the stellar half-mass radius, and divide it by
the corresponding time, to get the star formation rate for our sample of galaxies. We see a clear peak in the star
formation rate at the beginning of the merger, as expected both from observations and simulations (e.g.
\citealt{1978ApJ...219...46L,2007A&A...468...61D,2008AJ....135.1877E}). We also show the fraction of gas in the
baryonic component over time, and see that it starts quite high around the merger period (about 70\% of gas), and
then gradually decreases due to the star formation. After the star formation burst induced by the merger, the star
formation rate also decreases slowly over time, as the gas reservoir is depleted, but does not drop to zero until the
very end of the simulation. There is therefore continuous formation of stars in the remnant galaxy from the merger
until the present time.

Furthermore, if we look at the fraction of stars formed in the disc (defined with a thickness $|Z|<5$ kpc
and a cylindrical radius $R<50$ kpc) over time shown in Fig. \ref{SFdisc_samp}, we find that after the merger this fraction is very high,
over 90 \%, showing that most of the star formation occurs in the disc. Therefore our disc component is formed \textit{in situ}
from gradual and continuous star formation after the merger.

\begin{figure}
  \center
  \includegraphics[scale=0.35]{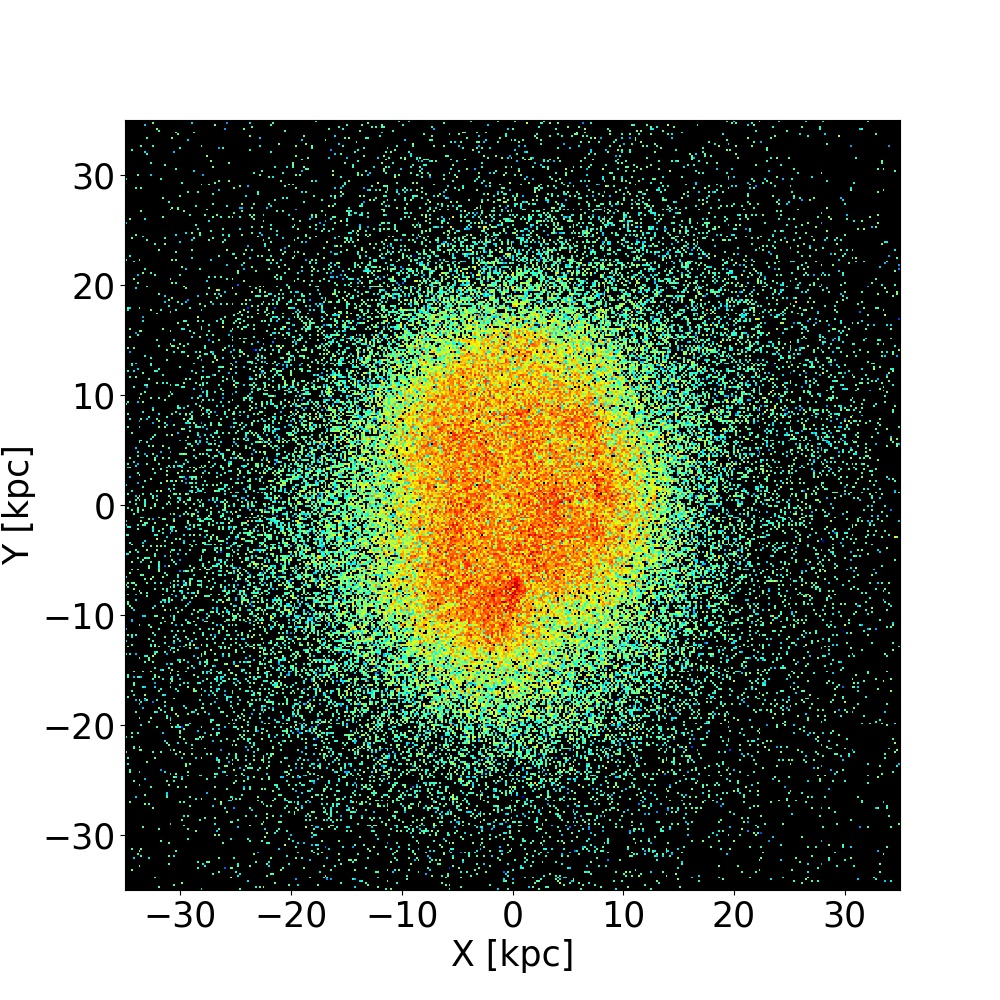}
  \includegraphics[scale=0.35]{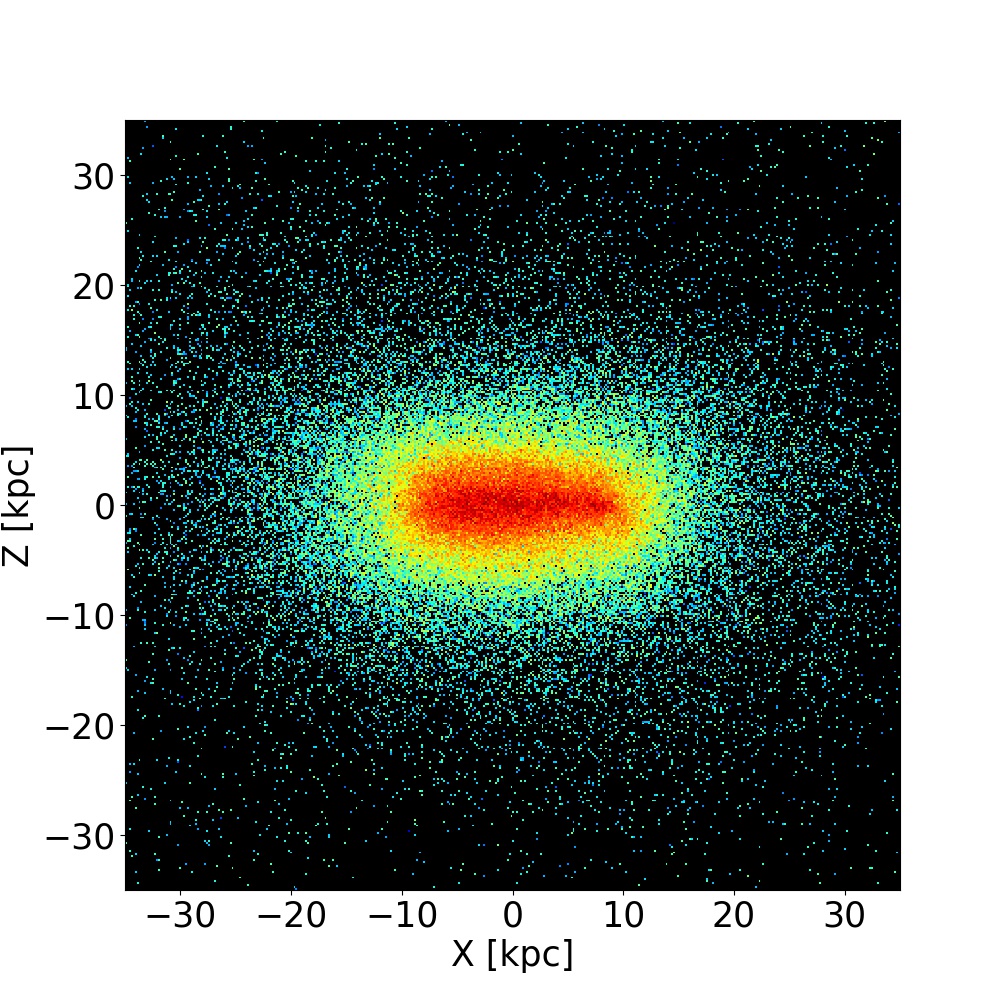}
\caption{Stellar density for Subhalo 16948 (fiducial case), seen at redshift $z=0$ face-on (top panel) and edge-on (bottom panel).}
  \label{16948_135}
\end{figure}

Using this sample we therefore get a general idea on how to form a disc from a major merger. The merger
destroys the galaxies and forms a rather spheroidal component out of the old stars, but after the merger star formation
creates a new young disc component, in good agreement with what was found in \cite{2017MNRAS.468..994P} for an isolated pair of galaxies. This is only possible in a merging system with enough gas, i.e. a wet merger, and
indeed in all our cases the galaxy right after the merger is gas-rich (around 70 \% of gas). This is consistent with
previous simulations of isolated galaxy mergers (e.g.
\citealt{2005ApJ...622L...9S,2009ApJ...691.1168H,2016ApJ...821...90A}).

\section{Fiducial example}
\label{fiduc}

We will now take one fiducial example of a galaxy from the disc sample described in the previous section and
investigate it in more detail, to better understand the formation of the disc and the role of gas in building it.

\subsection{Presentation of the case}

\begin{figure*}
  \center
  \includegraphics[scale=0.22]{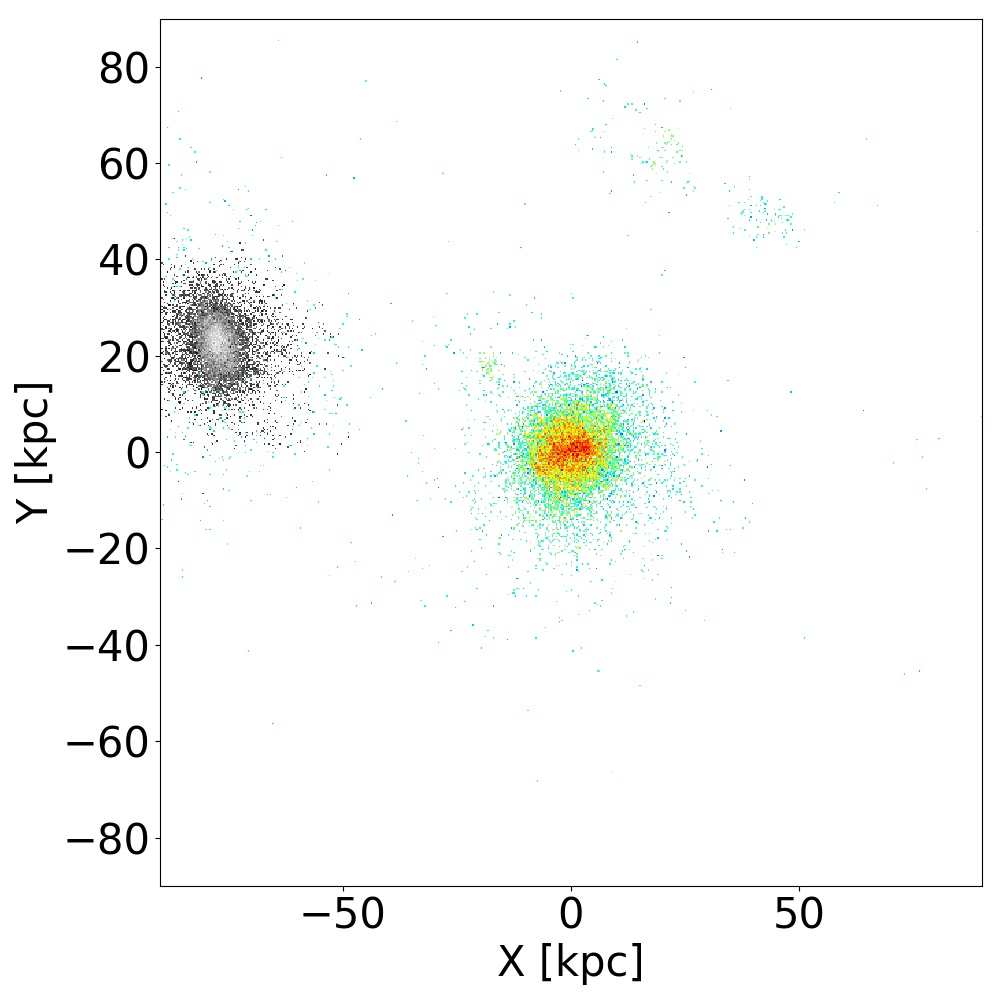}
  \includegraphics[scale=0.22]{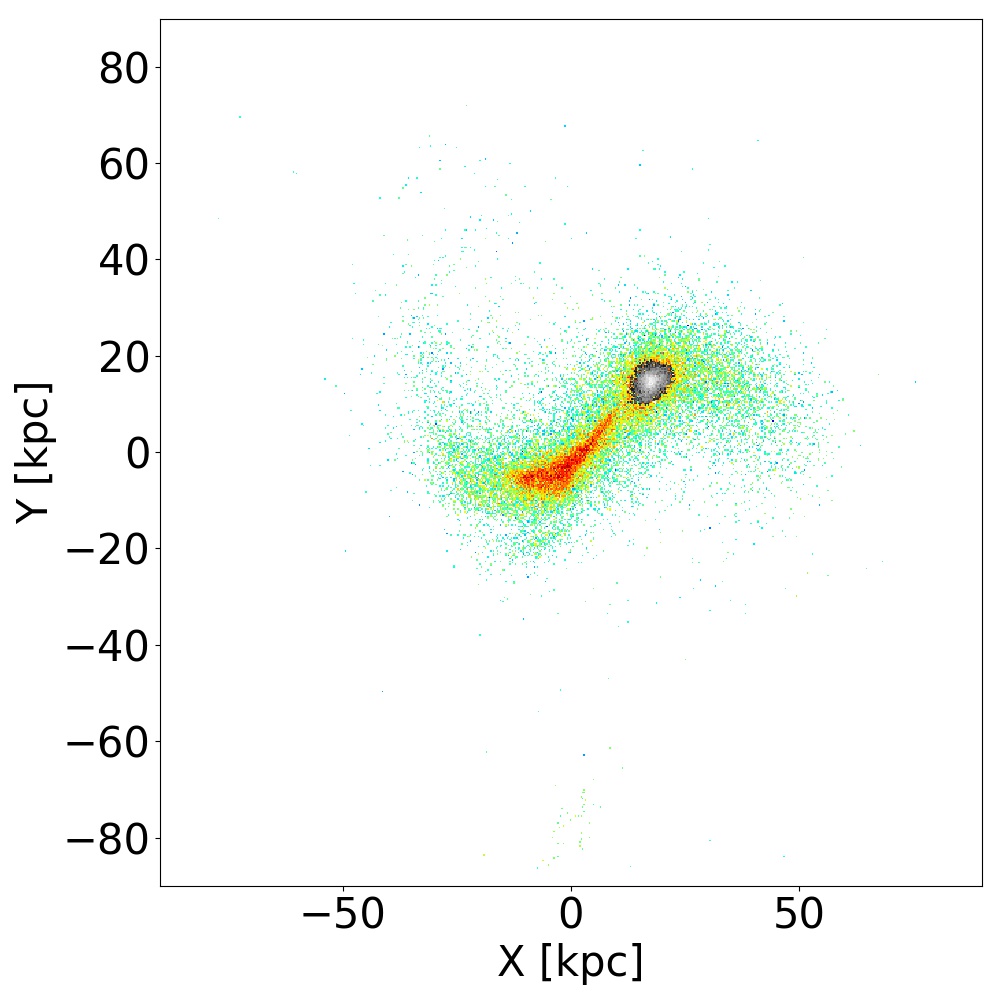}
  \includegraphics[scale=0.22]{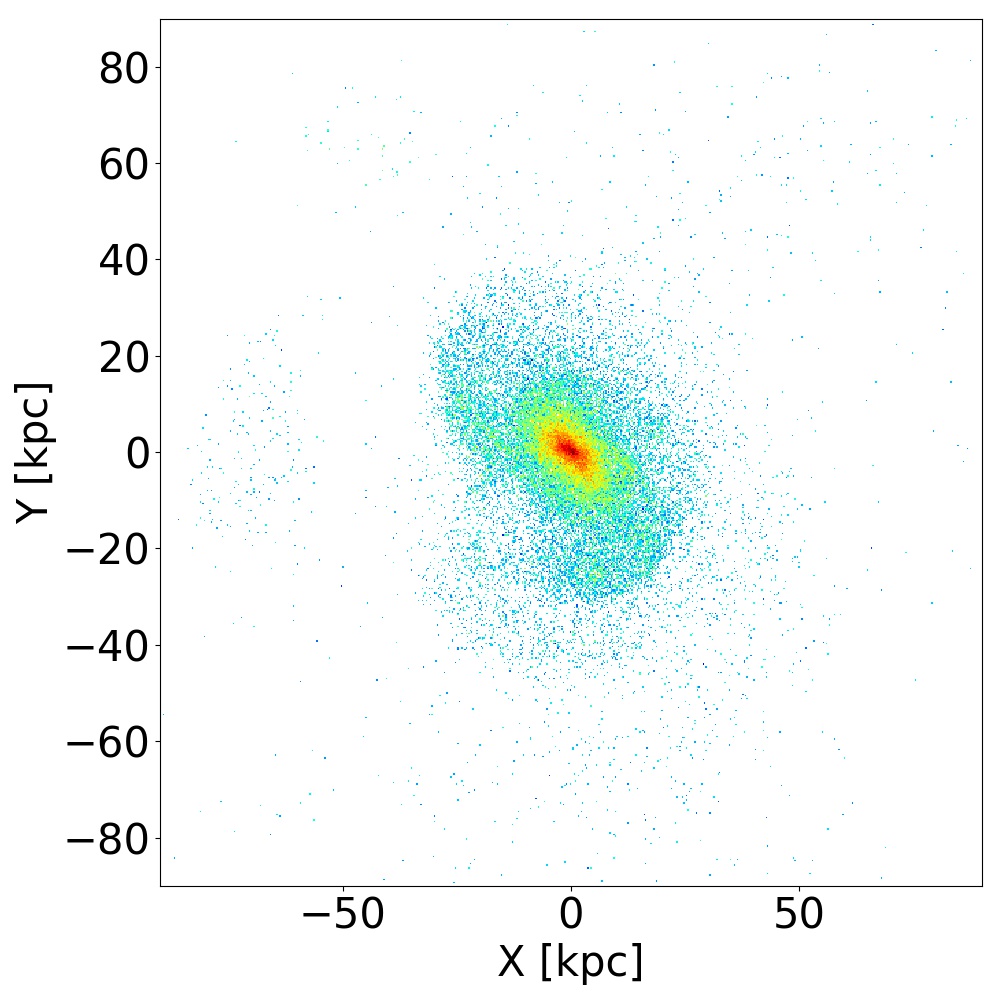}
\caption{Stellar density for Subhalo 16948 undergoing a major merger at redshift $z=1$, before (left panel), during
(middle panel) and after the merger (right panel). The primary galaxy is in color, and the secondary in greyscale. The
three corresponding times are 5.18, 5.61 and 6.23 Gyr after the beginning of the simulation.}
  \label{16948_merg}
\end{figure*}

The fiducial case we selected is the Illustris Subhalo number 16948 at snapshot 135 (displayed in Fig.
\ref{16948_135}), a disc galaxy at redshift $z=0$, with a total mass of $9.24 \times 10^{11} M_{\odot}$, a stellar mass
of $1.11 \times 10^{11} M_{\odot}$, a circular orbit fraction $f_{\epsilon} = 0.24$, and a flatness of 0.496 (as
defined in section \ref{res_disc}). The masses are computed as in section \ref{disc_sample_merg}. At redshift $z=1$, this
galaxy had a total mass of $3.5 \times 10^{11} M_{\odot}$ and underwent a major merger with a galaxy of total mass $2.6
\times 10^{11} M_{\odot}$. As the secondary galaxy is smaller and quickly distorted by the merger, to compute its total
mass we took the mass inside 5 times its stellar half mass radius.

We show in Fig. \ref{mass_16948} (top panel) the total masses of both merging galaxies as a function of time, where we
can see the secondary galaxy giving its total mass to the primary. We display the galaxy undergoing the merger in 3
snapshots (right before, during, and after the merger) in Fig. \ref{16948_merg}. As stated in section \ref{disc_glob},
the major merger completely destroys the stellar structures of both galaxies, and the disc is gradually formed only
after the merger (as will be shown in the following sections). The merger (as probed by eye) occurs between 5.47 Gyr
and 6.23 Gyr after the beginning of the simulation (redshift $z \sim$ 0.9 - 1.1).

\begin{figure}
  \center
\hspace{0.045cm}
  \includegraphics[scale=0.385]{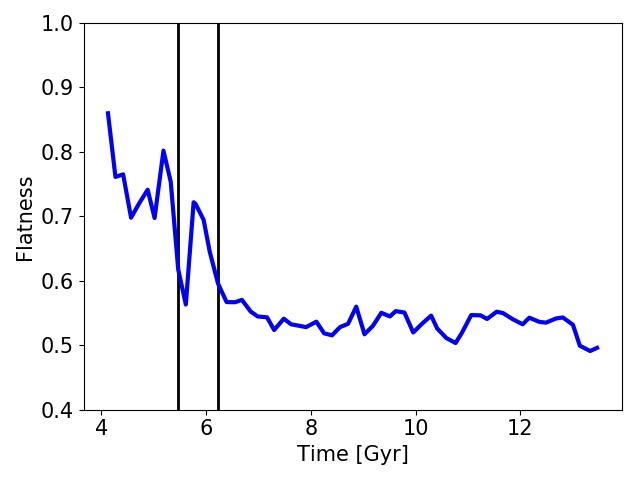} \\
\hspace{0.70cm}
  \includegraphics[scale=0.44]{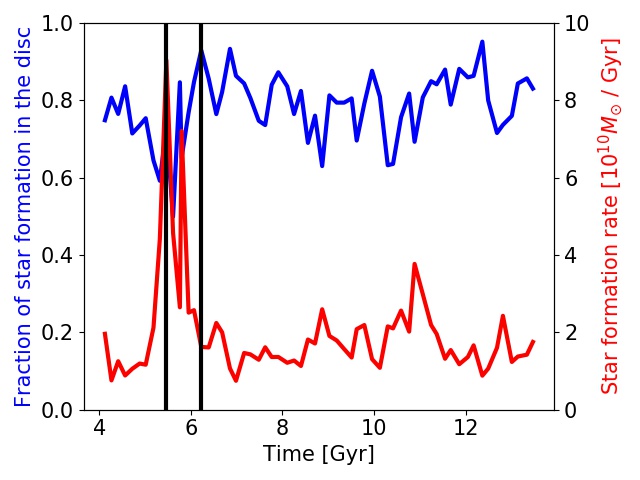}
  \includegraphics[scale=0.40]{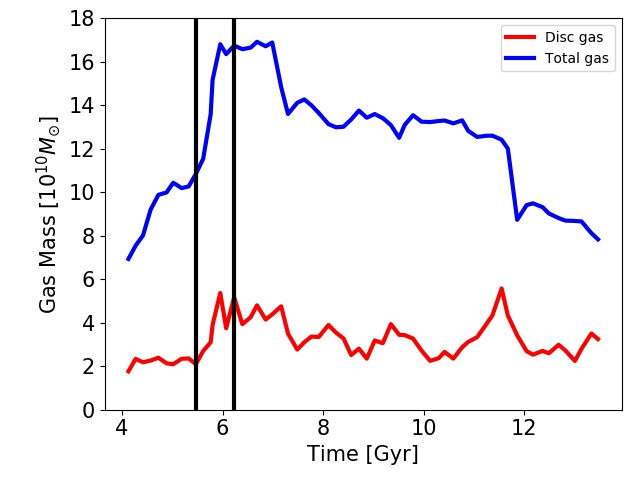}
\caption{Different quantities plotted as a function of time (= age of the Universe in Illustris) for the primary galaxy
in our fiducial case. Top panel: flatness as a function of time, as defined in section \ref{res_disc}. Middle panel:
star formation rate in red with the fraction of star formation occurring in the disc in blue, as a function of time.
Bottom panel: amount of gas in the disc in red, as well as in the whole galaxy in blue. The two black vertical lines
represent the merging period.}
  \label{time_16948}
\end{figure}

\subsection{Time evolution}
\label{16948_time}

To understand how the final disc galaxy was formed, we followed different quantities over time.

Looking at the stellar mass over time (Fig. \ref{mass_16948}, bottom panel), we see a sudden increase of the stellar
mass during the merger, due to the piling of stars in both galaxies as well as a star formation burst (see below). The
stellar mass then continuously increases up to redshift $z=0$, as new stars are formed.

We computed the stellar flatness of the galaxy, defined as in section \ref{res_disc}, and followed it through time.
Here again the flatness  before the merger is computed for the primary galaxy progenitor. We find that after the end of
the merger, the flatness drops and stays stable after a few snapshots (Fig. \ref{time_16948}, top panel). The stellar
disc thus seems to form rather quickly.

We also looked at the star formation over time in the galaxy in Fig. \ref{time_16948} (middle panel), and find that
after the star formation burst triggered by the merger, it remains rather constant, around $2 \times 10^{11}$
$M_{\odot}$ Gyr$^{-1}$. Furthermore, the star formation happens mostly in the disc after the merger ($\sim$80 \%, Fig.
\ref{time_16948}, middle panel), showing here also that \textit{in situ} continuous star formation is the main
mechanism building the disc.

\begin{figure}
  \includegraphics[scale=0.3]{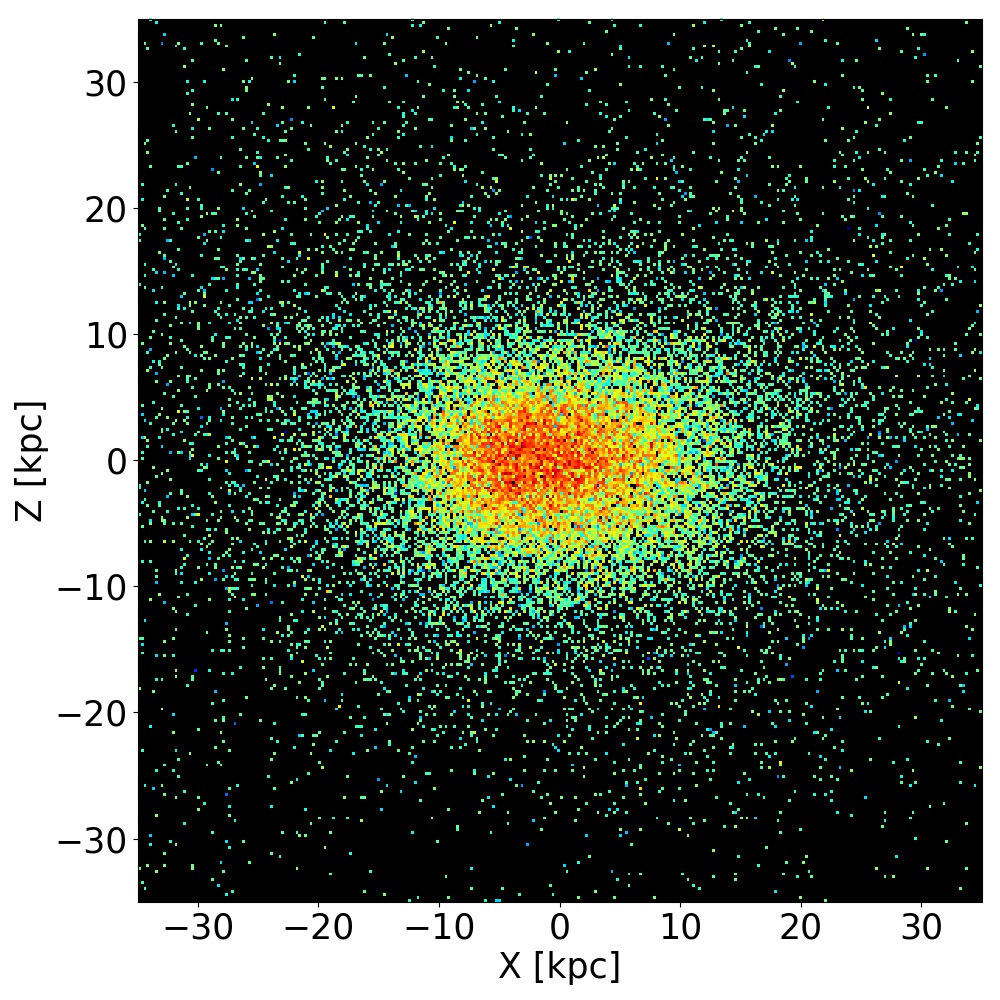}
   \includegraphics[scale=0.3]{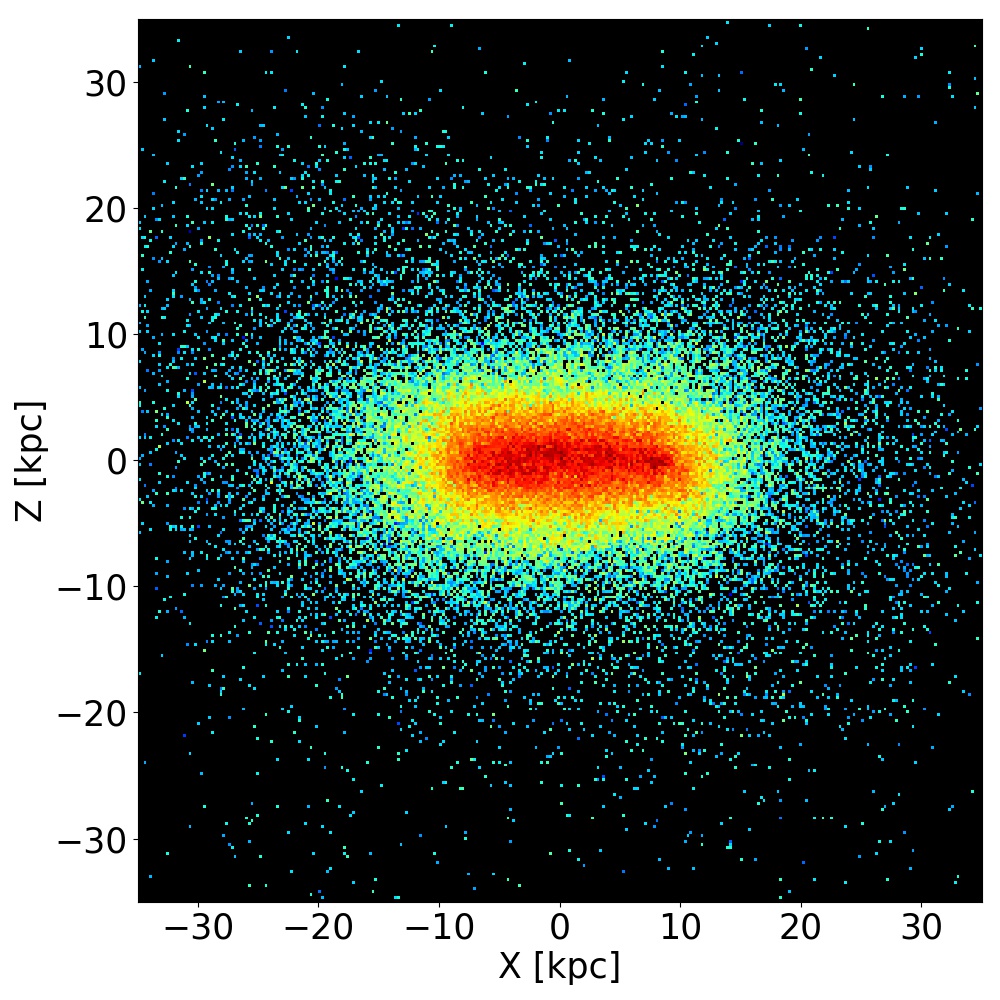}
\caption{Distribution of two stellar populations at redshift $z=0$ in our fiducial example (Subhalo 16948): stars formed before the
merger (top panel) and after the merger (bottom panel), in the galaxy seen edge-on.}
  \label{pop}
\end{figure}

To illustrate this result and the role of star formation, we decompose here again the stars into two populations: the
ones born before the merger (old stars), and the ones formed after (young stars). By displaying their distribution in
Fig. \ref{pop}, we see that the old stars form a rather ellipsoidal component, while the young ones form a clear disc,
in agreement with Fig. \ref{flat_ratio} and the results of section \ref{res_disc}. This is consistent with the results already found in \cite{2016ApJ...821...90A}.

Looking at the gas (Fig. \ref{time_16948}, bottom panel), we find that the quantity of gas decreases in the galaxy,
which can be explained by the star formation, as it is more important than the gas accretion rate. However, after a relatively sharp increase between 6 and 7 Gyr right after the merger, the quantity of gas in the disc then stays relatively stable over time, although
this is where we expect most of the star formation to happen and thus deplete the gas reservoir. For the amount of gas
to remain relatively constant, a continuous input of gas is needed in the disc, compensating the gas consumed in star formation.
We will investigate this point in the following section.

\subsection{The role of gas}
\label{16948_gas}

We showed in the previous section that most star formation happens in the disc component and there are very few stars
formed outside the disc. But where does this disc gas come from? Is this gas transformed into a disc component by the
merger, or does it come from slow accretion from the outer parts into the disc over time, consistently with the results in \cite{2017MNRAS.468..994P}? The amount of gas being
constant in the disc over time suggests the latter explanation, and we will here verify this hypothesis by following the gas movements.

We first start by looking at inflows and outflows of gas in the disc, to test our scenario that the gas is supplied from outside the disc over time. We compute the flux of gas going through the surfaces located 5 kpc above and under the disc, parallel to the disc with a 50 kpc cylindrical radius extent. To do this we take all particles located within a 0.5 kpc vertical distance from these surfaces, and divide them between particles going away from the disc (outflows) and particles going towards the disc (inflows). Although this is just an approximation (there might be particles entering or leaving the disc from the edges, without going through those two surfaces), this gives us an idea of what happens to the gaseous disc, and whether there is more gas entering or leaving it. We find that overall the gas inflow is dominant after the merger, with an average value of 2.1 $\times$ $10^{11}M_{\odot}Gyr^{-1}$, against only 1.3 $\times$ $10^{11}M_{\odot}Gyr^{-1}$ for outflows. It therefore seems that there is an important supply of gas to the disc, as our results in the previous section suggested. But where does this gas come from, and how big is its influence on the disc formation? To investigate this, we will now trace the gas directly to observe its movements.

Tracing the gas in time in Illustris is not straightforward, as contrary to stars or dark matter, gas cannot be
followed consistently from one snapshot to another in galaxies. An alternative solution is to use tracers. Tracers are
particles that are in the gaseous state at the beginning of the simulation in Illustris, and that are followed with a
fixed index throughout the whole simulation independently of what happens to them. This means that we can follow these
particles even if they transform into stars, or if they travel from one galaxy to another. In our case we will use them
to find out how the gas moves and forms stars in our fiducial example.

\begin{figure}
  \includegraphics[scale=0.3]{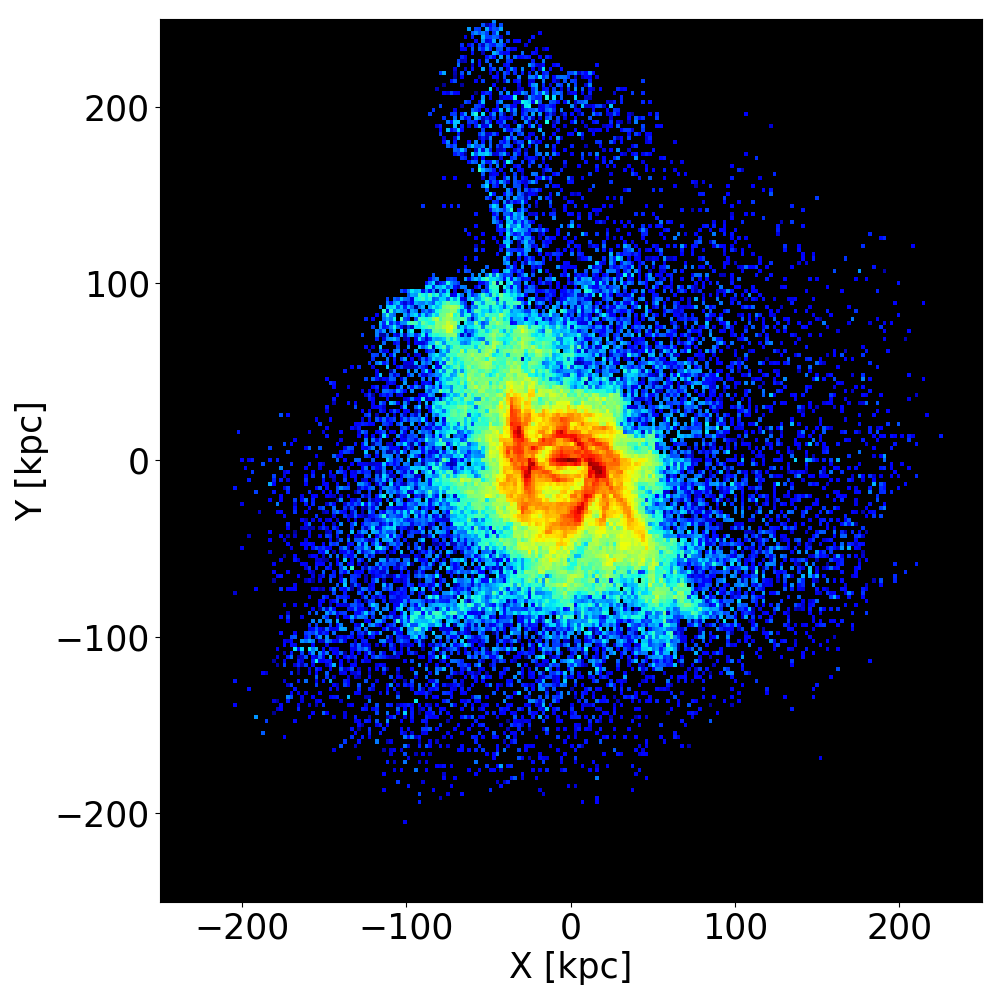}
  \includegraphics[scale=0.3]{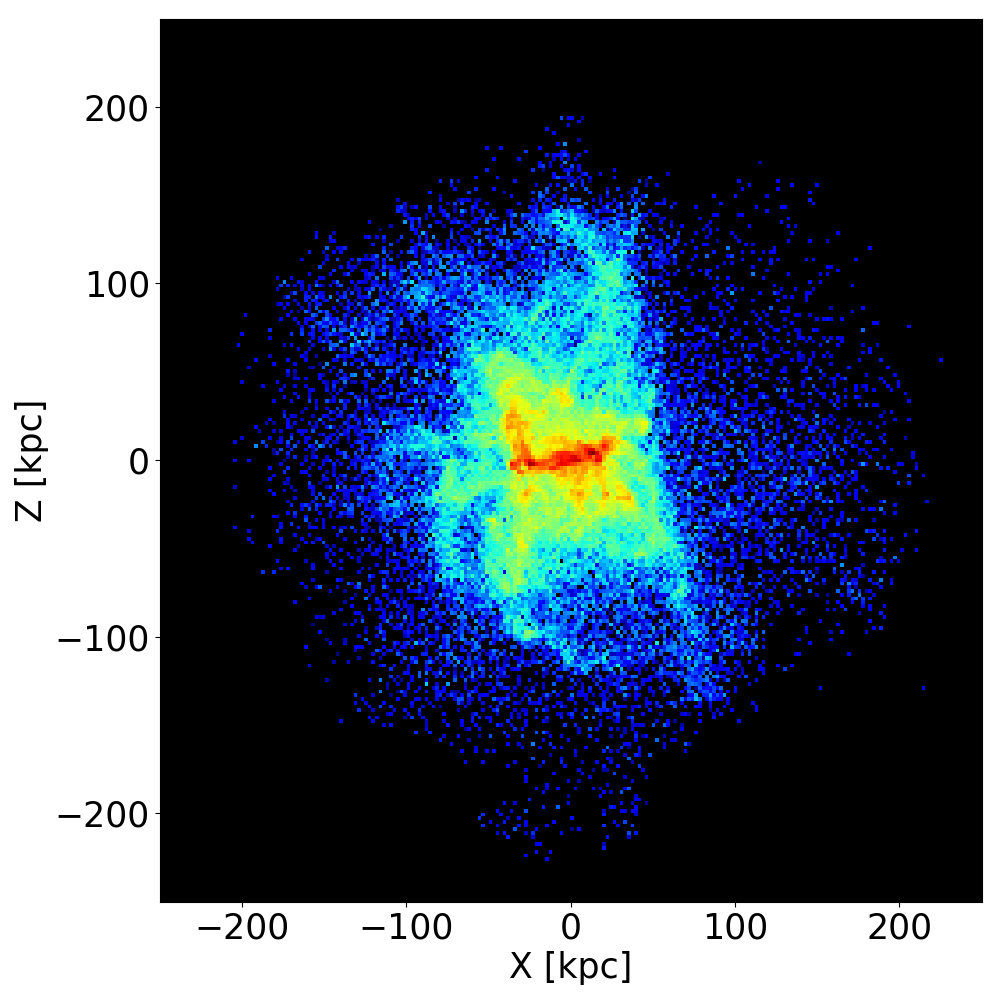}
\caption{Density distribution at snapshot 89 of the gas which forms stars by redshift $z=0$ in our fiducial case, in
face-on (top) and edge-on view (bottom).}
  \label{trac}
\end{figure}

We will therefore take the stellar tracers present in the disc of our galaxy at redshift $z=0$, and check where the
gas that formed them was at a given snapshot. This allows us to follow over time the gas that will end up as stars in the disc at the end of the simulation. Doing that for snapshot 89 (at the end of the merger), we find that a
significant fraction of this gas (87 \%) is in the halo of the galaxy, i.e. not in the disc. As in section
\ref{res_disc}, we define here the disc as the volume with $|Z|<5$ kpc and $R<50$ kpc. We display the
distribution of this gas, i.e. the gas that is forming the stellar disc by redshift $z=0$, in Fig. \ref{trac} face-on and edge-on. We
see how most of it is located out of the disc, but in its direct neighbourhood. We repeated this analysis at different
snapshots after the merger, and found the same result. It thus seems that most of the gas forming the future stellar
disc is initially in the halo part, surrounding the disc. The presence of this halo gas is not surprising, as it has been shown in observations that galaxies have hot gas surrounding their disc (e.g. \citealt{2015ApJ...800...14M}). In our case, this gas likely comes from the filament structures of our cosmological framework, as well as from the progenitors gas scattered away by the violence of the merger event. Since most of the star formation happens in the disc, it
means that this gas has necessarily been accreted into the disc before forming stars. Therefore, the disc is built from
gas present in the halo after the merger (as was already found in \citealt{2017MNRAS.468..994P} with the angular momentum), and gradually reaccreted in the disc over time. This is confirmed by the presence of gas inflows in the disc (see above), as well as the amount of gas staying constant over time in the disc
despite star formation (see section \ref{16948_time}); the gas transforming into stars is continuously replaced by gas accreted from the halo, allowing
constant refueling of the disc gas to form the stellar disc. This shows again the importance of the gas; the only way to
allow constant star formation and building of the disc (in the absence of external perturbations), is to use the gas
initially scattered in the halo by the merger, which thus has to be gas-rich.

\section{Formation of ellipticals from major mergers}
\label{ell_glob}

We have shown in the previous sections how a disc galaxy can arise from a major merger, however not all major mergers
produce disc galaxies. In this section we study the formation of elliptical galaxies from major mergers in Illustris,
as ellipticals are the most straightforward and expected outcome of major mergers, and we look for the differences
with the disc formation scenario investigated previously.

\subsection{Selection of the sample}
\label{ell_samp}

We follow the same procedure as in section \ref{disc_glob}, but this time selecting elliptical galaxies instead of disc
galaxies at redshift $z=0$. To be sure to select only pure ellipticals, we define our ellipticals as all galaxies with
again more than 60 000 stellar particles, but which have a flatness higher than 0.8, and less than 10\% of their stars
belonging dynamically to a disc (circularity parameter $\epsilon$ higher than 0.7). We then proceed exactly as in
section \ref{disc_sample_merg}, i.e. we trace back the history of these galaxies, and try to find major mergers. As before,
we dismiss all the cases where a significant minor merger occurs after the major merger, to focus on the major merger
as the main formation mechanism. This leaves us with 11 ellipticals formed from a major merger.

We furthermore dismiss four cases where the major merger has occurred at low redshift ($z<0.2$), as in these
cases not enough time has passed for a potential disc to form in the remnant, and we therefore cannot argue that the
major merger would not eventually produce a disc galaxy. This leaves us with a sample of 7 galaxies, whose total mass
distribution is displayed in the top panel of Fig. \ref{histg} (average total mass $(3.37 \pm 5.78) \times 10^{12} M_{\odot}$). In
this case, to compute the total mass, we did not use an exponential fit as in section \ref{disc_sample_merg}, as there is no
disc component in the galaxies. Instead we fit a de Vaucouleurs profile (\citealt{1948AnAp...11..247D}), and take the
particles inside 200 times the corresponding scale radius, as we probed by eye that this value contained most of the
mass. Note that we also tried to use 10 times the scale radius of a NFW (\citealt{1997ApJ...490..493N}) profile fitted on the dark matter distribution of these galaxies, with similar results for the derived total masses. The list and properties of those 7 galaxies is given in the Appendix (Table \ref{table1}).

As in section \ref{disc_sample_merg}, we checked whether our galaxies were indeed realistic ellipticals, by looking at their rotation and their surface density profiles (Fig. \ref{histg}, middle panel). We find that the galaxies of the elliptical sample have indeed very low ratios of rotation to dispersion ($0.053 \pm 0.050$, see individual values in the Appendix), much lower than our disc sample ($0.55\pm 0.15$), showing that they are dispersion dominated. Furthermore, by fitting their surface density profiles with a S\'ersic function, we find high S\'ersic indexes, around 2 or higher (see Appendix for the values), and with a distribution tending to larger values than for our disc sample (Fig. \ref{histg}, bottom panel). This confirms that our elliptical sample is indeed composed of early type galaxies, and not disc galaxies.

\subsection{Results}
\label{res_ell}

Those 7 cases represent major mergers where no consistent disc has formed in the remnant galaxy. While in section
\ref{res_disc} we analyzed why and how a disc was formed in the remnant, here we will try to understand why no disc has
formed.

\begin{figure}
  \includegraphics[scale=0.5]{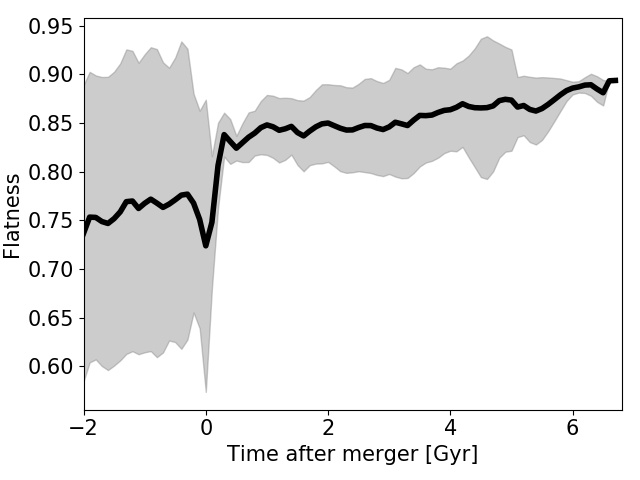}
\caption{Flatness (as defined in section \ref{res_disc}) as a function of time (or snapshots) for the sample of
elliptical galaxies formed from major mergers. As in Fig. \ref{flat_samp}, \ref{sfr_samp} and \ref{SFdisc_samp}, the
snapshots are scaled with respect to the time of the merger for each galaxy.}
\label{flat_ell}
\end{figure}

We first plot the flatness as a function of time for our sample (Fig. \ref{flat_ell}), and see that the flatness
remains high and constant after the merger, which shows indeed that no clear disc has formed in the remnant.

\begin{figure}
  \includegraphics[scale=0.5]{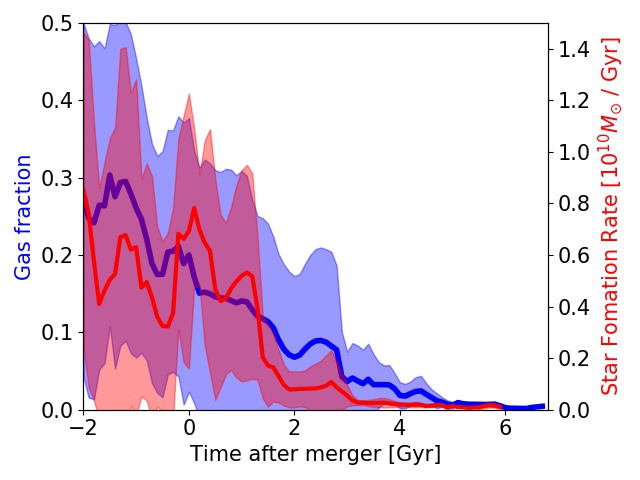}
\caption{Same as Fig. \ref{sfr_samp}, but this time for our sample of elliptical galaxies.}
  \label{sfr_ell}
\end{figure}

Looking at the gas content, we find that in all our cases there is little gas in the remnant galaxy. We characterize
this result by plotting the amount of gas as a function of time, as well as the star formation rate, for our sample
(Fig. \ref{sfr_ell}). We see that the amount of gas in the galaxy is much smaller (around 10 \% of the baryonic
matter) at the merger time than in the disc sample of section \ref{res_disc}, and drops quickly after the merger. The star formation rate
is also small, as expected, and drops quickly to nearly zero. Therefore, after the merger the absence of gas seems to
prevent the consistent formation of a young stellar component.

\begin{figure}
  \includegraphics[scale=0.3]{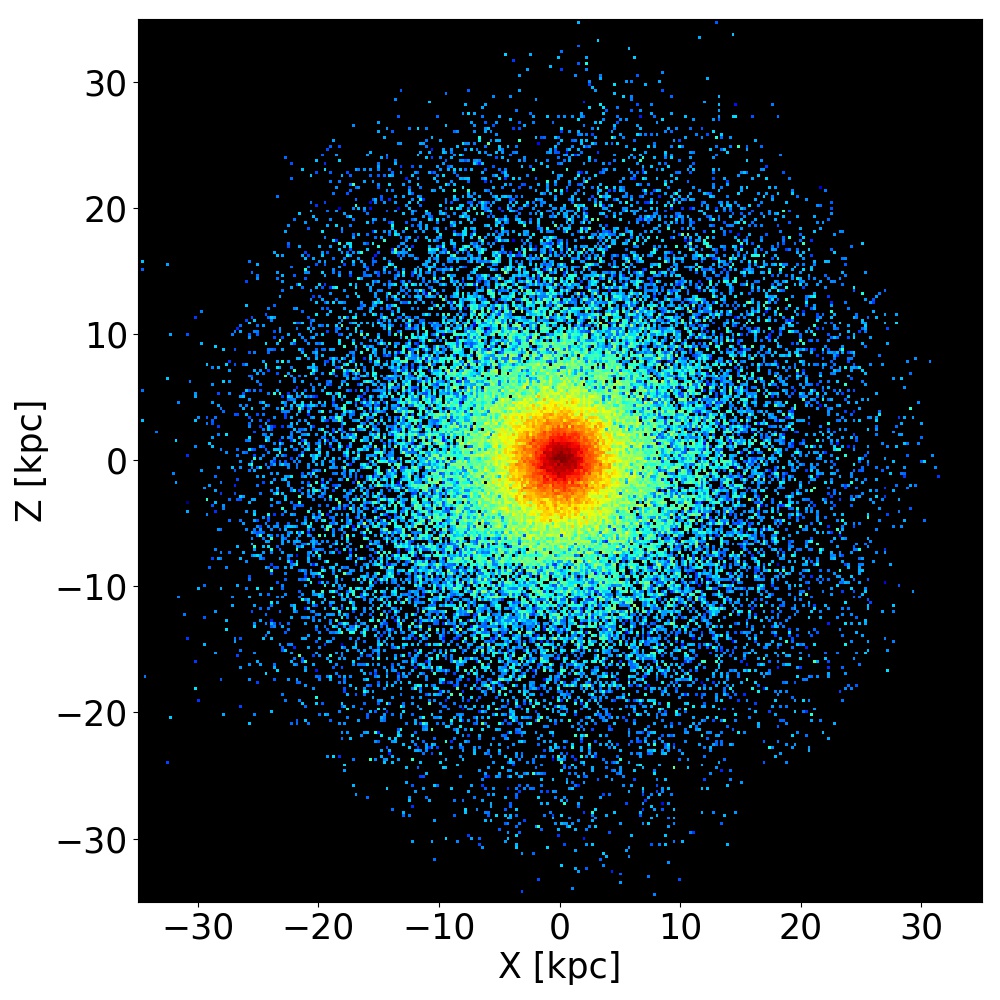}
  \includegraphics[scale=0.3]{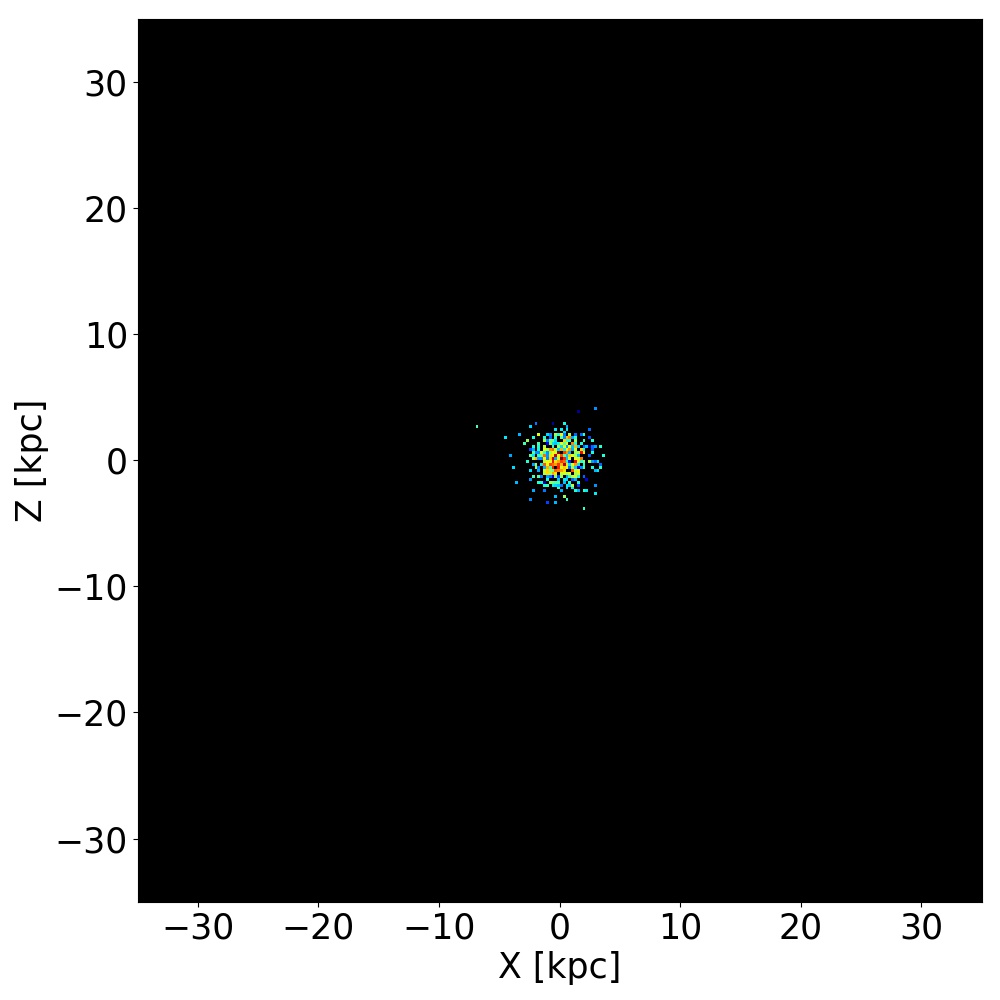}
\caption{Division into stellar populations of different ages (similar to Fig. \ref{pop}), for an elliptical galaxy of
our sample (Subhalo 59397 at redshift $z=0$), seen edge-on (Z being the rotation axis of the galaxy). There is no trace of a disc component in any population, the stars formed
after the merger are few and concentrated in the central part of the galaxy.}
  \label{popell_ex}
\end{figure}

If we divide the stellar populations into several age bins as in Fig. \ref{pop}, we find no trace of a disc component
in the younger populations, except for one galaxy, Subhalo number 16960 (number at redshift $z=0$). This galaxy has a
small thin young disc, but of a very low mass (with the young population containing $\sim$ 15 \% of the total stellar
mass), so that it is small on the galactic scale and the galaxy remains dominated by its main, old spheroidal
component. In all other cases the small young population is concentrated in the central part (see an example in Fig.
\ref{popell_ex} for Subhalo number 59397), where most of the gas is expected to pile up due to the merger (\citealt{1996ApJ...471..115B}). We find
overall that the younger population (formed after the merger) contributes only 1.5 $\pm$ 2.2 \% of the stellar mass in
the galaxies in our sample (excluding Subhalo 16960).

Therefore, these 7 major mergers producing elliptical galaxies correspond to cases with little gas and/or little star
formation, which prevents the formation of a new extended disc in the remnant galaxy from accreted gas. The stars in
these galaxies are predominantly old (formed before the merger), and have been shuffled into a spheroidal component by
the violent relaxation of the merger, which created the dominant spheroidal component characteristic of elliptical
galaxies. This is consistent with the work of \cite{1977egsp.conf..401T}, \cite{2013ApJ...778...61T} and
\cite{2014MNRAS.444.3357N} on major mergers. We thus conclude that major mergers can produce both elliptical and disc
galaxies, depending mainly on the amount of gas present in the galaxy right after the merger.

\vspace{15mm}

\section{Discussion}
\label{discuss}

This work constitutes the continuation of the previous work of \cite{2016ApJ...821...90A, 2017A&A...600A..25R, 2017MNRAS.468..994P, 2017MNRAS.467L..46A, 2018IAUS..334...65A}, where an isolated system of two gas-rich protogalaxies on a merging orbit was simulated, and it was shown that the remnant is a spiral galaxy. Those simulations were done using the N-body/SPH code GADGET3 (\citealt{2002MNRAS.333..649S,2005MNRAS.364.1105S}), with 5.5 million particles representing dark matter, gas, and stars. The two protogalaxies are initially only constituted of dark matter and a hot gaseous halo, and both form small stellar discs, which are then destroyed by the merger to form a bulge component. Gas from the halo then plays a fundamental role after the merger, being accreted to form a new disc in the remnant, giving it its angular momentum, and \textit{via} star formation allowing the creation of a stellar disc with well-defined spiral arms and a bar. The results presented in this paper are therefore in very good agreement with the latter, as we also show how the gas from the halo is accreted to form a stellar disc after the merger. Despite a necessarily lower resolution, we extended this previous work by showing that those results are also valid in the more realistic frame of a cosmological context. 

In section \ref{disc_glob}, we took all major mergers happening from $z=0$ up to $z=1.5$ for our disc galaxies sample,
and discarded all major mergers happening earlier. The choice of $z=1.5$ as a limiting value is justified by the fact
that at earlier times galaxy mergers are much less clear, as galaxies are smaller, there are many multiple mergers
happening (more than two galaxies merging at the same time), and the evolution of early galaxies is generally harder to
track.

We tried to look at the flatness of the gas component besides the stellar one, to see if the accretion was creating
a gaseous disc. However we did not see any clear trend, the gas is predominantly in a spheroidal component outside the
disc (see section \ref{16948_gas}) so that its flatness remains close to 1 also during episodes of gas
accretion. Furthermore the accreted gas is quickly transformed into stars, so that the creation of a real gaseous disc
is not clearly visible. 

\cite{2017MNRAS.470.3946S} also analyzed major mergers creating discs in Illustris, but for a much smaller sample of
four galaxies. They performed zoom-in re-simulations of those galaxies (\citealt{2016MNRAS.462.2418S}), allowing
better resolution, to investigate the quenching and the effect of feedback on it. They proved that it is possible in
Illustris to form star forming disc galaxies after a major merger, depending on the strength of AGN feedback. Our
results confirm that Illustris can form disc galaxies after a major merger, and go further, as our larger sample allows
better statistics on the properties of those galaxies over time. While \cite{2017MNRAS.470.3946S} focused on the
quenching and feedback, we focussed the formation of the disc over time, using its flatness, and tracing the
gas forming it. This allowed us to show how important the amount of gas is in this process, as it is accreted from the
halo after the merger. Furthermore, we added the study of elliptical galaxies from dry mergers to complete our analysis
and show the importance of the amount of gas in the disc formation process.

Comparing our sample to \cite{2017MNRAS.470.3946S}, we looked for mergers happening in a larger redshift range (up to z=1.5) and took a weaker constraint on
the mass ratios ($M_2/M_1>4/5$ in their sample, while we have 1/4), so that our results are valid for a broader range of merger parameters, at higher redshift and are not limited to
galaxies of very similar masses. Furthermore, while \cite{2017MNRAS.470.3946S} just made sure that there was no other
massive galaxy (with a mass ratio higher than 1/3) at $z=0.2$ and 0.5 around their studied galaxies, we paid extra
attention to excluding any minor mergers from the disc formation process, by checking at every snapshot following the
merger that there is no encounter involving a galaxy with a mass ratio higher than 1/20. This ensures that the disc is
formed from the major merger only, without the contribution of any other external event, and therefore makes our claim stronger. 

\cite{2017MNRAS.467.3083R} also established a link between the amount of gas and the morphology of the remnant galaxy in
Illustris, but using a different approach. They used the $\kappa_{rot}$ parameter (\citealt{2012MNRAS.423.1544S}) based
on the kinetic energy, to account for the amount of rotational support of the stellar component. They took a big
sample (18000) of galaxies and found a correlation between the amount of gas in the merger and  $\kappa_{rot}$, with dry
mergers producing galaxies with less rotational support. Our results therefore support this claim, as one can assume
that disc galaxies usually have a high $\kappa_{rot}$, while ellipticals have a low value of this parameter.
However, the approach and aim of our paper is very different from those of \cite{2017MNRAS.467.3083R}. While they
used their high number of galaxies to extract statistics on the outcome of mergers, we decided to focus on a smaller
sample, although still significant enough to derive statistics, and have more control over the types of galaxies and
mergers occurring. We restrict ourselves to clear disc galaxies and ellipticals, examine our galaxies individually,
focus on major mergers only, and make sure that the major merger is the only significant event able to impact the
galaxy morphology. Besides, we support our claim by looking in detail at the formation of the disc, to understand
how it happens, from the moment of the merger until the present time. Our analysis is therefore more targeted and focused
on the effect of major mergers, with in-depth study of the building of the disc and the role of gas. 

Note that in this paper we do not claim that all wet major mergers produce disc galaxies, there are many factors that
can prevent the formation of a disc or destroy an existing one, such as minor mergers (see below), tidal interactions or gas removal after the merger. We only study cases of
galaxies that we know end up as disc galaxies and had a major merger in their history, to show that it is indeed
possible, even in a cosmological context, to create disc galaxies from major mergers, and to study in details their properties and the mechanisms
involved. Similarly, we do not claim that dry mergers always produce ellipticals (and indeed previous collisionless simulations have shown that dry mergers do not necessarily lead to galaxies resembling present-day ellipticals, see e.g. \citealt{2006MNRAS.369..625N}), we selected our sample to have only
elliptical galaxies at redshift $z=0$.

Minor mergers are not studied in this work, but could potentially have a significant impact on the remnant galaxy (or
on any galaxy in general). A violent collision with a relatively big satellite (for example with a mass ratio around
1/5) could disturb a disc structure that has formed from a wet major merger, potentially even destroying the disc
(\citealt{2008ApJ...688..757H}). Therefore we do not expect every major merger to lead to a disc galaxy. Furthermore,
external events affecting the amount of gas, such as gas inflows or outflows, will obviously affect the formation of
the disc, possibly triggering it or preventing it. 

We have only 7 cases of elliptical galaxies in our sample in section \ref{ell_glob}, which might appear a small
number, and is indeed smaller than the sample of disc galaxies in section \ref{disc_glob}. This does not reflect the frequency of ellipticals and discs in Illustris at redshift $z=0$, which is in good agreement with observations (see \citealt{2015MNRAS.454.1886S}), and therefore the difference in the size of our two samples must be due to the way we selected them. Besides our restrictive
constraints (major merger happening between redshift $z=0.2$ and $z=1.5$, mass range, etc.), we explain this by the
fact that we removed all the cases where a minor merger (with a mass ratio higher than 5 \%) happens after the major
merger. Since ellipticals are often very massive galaxies, they tend to accrete numerous satellites around them, so
that minor mergers are very frequent, and we lose many cases by removing the galaxies which had them.  However for our
analysis it is necessary to isolate the process of major merger, and therefore to discard the cases where a significant
minor merger happens in the history of the galaxy.

The constraint of having no significant minor mergers is a strong one, and might not reflect the evolution of most cluster galaxies, for which it is common to have satellites with 5 to 10\% of their mass to be captured and accreted (\citealt{2013MNRAS.428..641O, 2014MNRAS.444.3357N}). However the point of this paper is to prove that a major merger alone can lead to a disc galaxy in a cosmological context, and to understand the mechanisms involved, more than trying to characterize the typical evolution of a galaxy after a major merger. Nevertheless, to check that the scenario of discs formed from a major merger still works including more significant minor mergers, we tried to take a weaker constraint, by changing the maximum mass of satellites to be accreted from 5\% to 10\% of the galaxy's total mass. As expected, this considerably increases the number of cases, as in this way we find over 200 disc galaxies formed from a major merger in Illustris, with the same mass constraints (see section \ref{disc_samp} and \ref{disc_sample_merg}). This confirms that it is also possible to create a disc galaxy from a major merger with subsequent minor mergers, although in those cases it becomes difficult to disentangle the effect of the major merger from the minor mergers on the disc formation.

In this work, we took two extreme samples at redshift $z=0$, the disc galaxies which are clearly dominated by their disc
component, and the elliptical galaxies having almost no disc component (based on the flatness and circularity of the
stars, see section \ref{disc_samp} and \ref{ell_samp}, but also on the ratio of rotation to dispersion and on the S\'ersic index of the galaxy, see middle and lower pannels of Fig. \ref{histg}). These cases are characterized by a lot of star formation in the
former case, and almost no star formation in the latter. However there are obviously many intermediate cases where the
presence of some gas in the remnant galaxy may lead to the formation of a small disc in the remnant, but not big
enough to dominate the galaxy. We mentioned such a case as one out of 7 cases in our elliptical sample in section
\ref{res_ell}, Subhalo 16960. This galaxy has a gas fraction a bit higher than the other cases at the moment of
the merger (around 20 \%, while the mean of the sample is around 12 \%). It also shows a slightly higher star
formation rate than other galaxies for a short amount of time (about 1 Gyr), which allows the formation of this low-mass disc.

\section{Summary and Conclusions}
\label{ccl}

In this paper, we studied how disc galaxies can form from major mergers in the Illustris simulation. We selected a
sample of 38 disc galaxies at redshift $z=0$ with a stellar mass higher than 5 $\times$ $10^{10}M_{\odot}$, that
experienced a major merger in their history after redshift 1.5, with no significant other merger (minor or major)
following. We found that at times after the merger the stellar component evolves gradually toward
a disc component.  The gas fraction is quite high after the merger (around 70 \%), which allows a lot of star formation
in the remnant galaxy even long after the merger. Most of this star formation occurs in the disc region, and
contributes to building the stellar disc. Indeed, the disc is composed predominantly of young stars, as the older
population (formed before the merger) has a much more spheroidal distribution. The disc is therefore formed gradually
by star formation after the merger, in good agreement with what was found in controlled simulations (\citealt{2016ApJ...821...90A,2017MNRAS.468..994P}).

We then focused on one fiducial case, Subhalo 16948 (index at redshift $z=0$), to study how the gas contributes to the
creation of the new disc. The amount of gas in the disc is stable over time despite star formation, which means there
must be an inflow of gas into the disc, which we showed by computing the flux of gas towards the disc plane. We used Illustris tracers to trace back where the gas that formed the young disc
population came from, and found that the star forming gas right after the merger is located mostly (87 \%) in the halo
of the remnant galaxy, surrounding the disc. Since most of the stars are formed in the disc region, this gas is
accreted gradually into the disc, where it cools down and builds the stellar disc by star formation, as already found in controlled simulations (\citealt{2017MNRAS.468..994P}).

Finally, we explored major mergers leading to elliptical galaxies in Illustris, by forming a sample of 7 elliptical
galaxies having experienced a major merger in their history. These elliptical galaxies were selected to also have a
stellar mass higher than 5 $\times$ $10^{10}M_{\odot}$ at redshift $z=0$, and a merger between $z=0.2$ and $z=1.5$. In
this case there is no flattening of the stellar component after the merger, i.e. no disc component is formed. We found
that these mergers are dry, i.e. they have little gas and very little star formation follows the merger. Therefore no
new stellar disc can be created in the remnant from accreted gas, and the old stellar population scattered into a
spheroidal component dominates the galaxy, which becomes an elliptical.

The gas is therefore essential to determine the outcome of a major merger, with dry mergers leading often to ellipticals,
and wet mergers creating disc galaxies. While these results were previously found in simulations of idealized mergers,
we showed here that this scenario also works in the more general and realistic frame of a cosmological context, with no
restricting conditions applied to the mergers or the environment.

\section*{Acknowledgments}

We are grateful to our referee, Brant Robertson, for suggestions that helped to improve the paper. We thank Ivana Ebrova, Grzegorz Gajda, Klaudia Kowalczyk, and Marcin Semczuk for useful discussions and
comments. This work was supported by the Polish National Science Centre under grants 2013/10/A/ST9/00023 and
2018/28/C/ST9/00443.

\label{lastpage}

\bibliographystyle{mnras}
\bibliography{biblio}

\begin{thebibliography}{}
\makeatletter
\relax
\def\mn@urlcharsother{\let\do\@makeother \do\$\do\&\do\#\do\^\do\_\do\%\do\~}
\def\mn@doi{\begingroup\mn@urlcharsother \@ifnextchar [ {\mn@doi@}
  {\mn@doi@[]}}
\def\mn@doi@[#1]#2{\def\@tempa{#1}\ifx\@tempa\@empty \href
  {http://dx.doi.org/#2} {doi:#2}\else \href {http://dx.doi.org/#2} {#1}\fi
  \endgroup}
\def\mn@eprint#1#2{\mn@eprint@#1:#2::\@nil}
\def\mn@eprint@arXiv#1{\href {http://arxiv.org/abs/#1} {{\tt arXiv:#1}}}
\def\mn@eprint@dblp#1{\href {http://dblp.uni-trier.de/rec/bibtex/#1.xml}
  {dblp:#1}}
\def\mn@eprint@#1:#2:#3:#4\@nil{\def\@tempa {#1}\def\@tempb {#2}\def\@tempc
  {#3}\ifx \@tempc \@empty \let \@tempc \@tempb \let \@tempb \@tempa \fi \ifx
  \@tempb \@empty \def\@tempb {arXiv}\fi \@ifundefined
  {mn@eprint@\@tempb}{\@tempb:\@tempc}{\expandafter \expandafter \csname
  mn@eprint@\@tempb\endcsname \expandafter{\@tempc}}}

\bibitem[\protect\citeauthoryear{{Athanassoula}}{{Athanassoula}}{2018}]{2018IAUS..334...65A}
{Athanassoula} E.,  2018, in {Chiappini} C.,  {Minchev} I.,  {Starkenburg} E.,
   {Valentini} M.,  eds,  IAU Symposium Vol. 334, Rediscovering Our Galaxy. pp
  65--72 (\mn@eprint {arXiv} {1801.07720}), \mn@doi{10.1017/S1743921317008778}

\bibitem[\protect\citeauthoryear{{Athanassoula}, {Rodionov}, {Peschken}  \&
  {Lambert}}{{Athanassoula} et~al.}{2016}]{2016ApJ...821...90A}
{Athanassoula} E.,  {Rodionov} S.~A.,  {Peschken} N.,   {Lambert} J.~C.,  2016,
  \mn@doi [\apj] {10.3847/0004-637X/821/2/90}, \href
  {https://ui.adsabs.harvard.edu/abs/2016ApJ...821...90A} {821, 90}

\bibitem[\protect\citeauthoryear{{Athanassoula}, {Rodionov}  \&
  {Prantzos}}{{Athanassoula} et~al.}{2017}]{2017MNRAS.467L..46A}
{Athanassoula} E.,  {Rodionov} S.~A.,   {Prantzos} N.,  2017, \mn@doi [\mnras]
  {10.1093/mnrasl/slw255}, \href
  {https://ui.adsabs.harvard.edu/abs/2017MNRAS.467L..46A} {467, L46}

\bibitem[\protect\citeauthoryear{{Barnes} \& {Hernquist}}{{Barnes} \&
  {Hernquist}}{1992}]{1992ARA&A..30..705B}
{Barnes} J.~E.,  {Hernquist} L.,  1992, \mn@doi [\araa]
  {10.1146/annurev.aa.30.090192.003421}, \href
  {https://ui.adsabs.harvard.edu/abs/1992ARA&A..30..705B} {30, 705}

\bibitem[\protect\citeauthoryear{{Barnes} \& {Hernquist}}{{Barnes} \&
  {Hernquist}}{1996}]{1996ApJ...471..115B}
{Barnes} J.~E.,  {Hernquist} L.,  1996, \mn@doi [\apj] {10.1086/177957}, \href
  {https://ui.adsabs.harvard.edu/abs/1996ApJ...471..115B} {471, 115}

\bibitem[\protect\citeauthoryear{{Borlaff} et~al.,}{{Borlaff}
  et~al.}{2014}]{2014A&A...570A.103B}
{Borlaff} A.,  et~al., 2014, \mn@doi [\aap] {10.1051/0004-6361/201424299},
  \href {https://ui.adsabs.harvard.edu/abs/2014A&A...570A.103B} {570, A103}

\bibitem[\protect\citeauthoryear{{Deeley} et~al.,}{{Deeley}
  et~al.}{2017}]{2017MNRAS.467.3934D}
{Deeley} S.,  et~al., 2017, \mn@doi [\mnras] {10.1093/mnras/stx441}, \href
  {https://ui.adsabs.harvard.edu/abs/2017MNRAS.467.3934D} {467, 3934}

\bibitem[\protect\citeauthoryear{{Di Matteo}, {Combes}, {Melchior}  \&
  {Semelin}}{{Di Matteo} et~al.}{2007}]{2007A&A...468...61D}
{Di Matteo} P.,  {Combes} F.,  {Melchior} A.~L.,   {Semelin} B.,  2007, \mn@doi
  [\aap] {10.1051/0004-6361:20066959}, \href
  {https://ui.adsabs.harvard.edu/abs/2007A&A...468...61D} {468, 61}

\bibitem[\protect\citeauthoryear{{Drory} \& {Fisher}}{{Drory} \&
  {Fisher}}{2007}]{2007ApJ...664..640D}
{Drory} N.,  {Fisher} D.~B.,  2007, \mn@doi [\apj] {10.1086/519441}, \href
  {https://ui.adsabs.harvard.edu/abs/2007ApJ...664..640D} {664, 640}

\bibitem[\protect\citeauthoryear{{Eliche-Moral}, {Rodr{\'\i}guez-P{\'e}rez},
  {Borlaff}, {Querejeta}  \& {Tapia}}{{Eliche-Moral}
  et~al.}{2018}]{2018A&A...617A.113E}
{Eliche-Moral} M.~C.,  {Rodr{\'\i}guez-P{\'e}rez} C.,  {Borlaff} A.,
  {Querejeta} M.,   {Tapia} T.,  2018, \mn@doi [\aap]
  {10.1051/0004-6361/201832911}, \href
  {https://ui.adsabs.harvard.edu/abs/2018A&A...617A.113E} {617, A113}

\bibitem[\protect\citeauthoryear{{Ellison}, {Patton}, {Simard}  \&
  {McConnachie}}{{Ellison} et~al.}{2008}]{2008AJ....135.1877E}
{Ellison} S.~L.,  {Patton} D.~R.,  {Simard} L.,   {McConnachie} A.~W.,  2008,
  \mn@doi [\aj] {10.1088/0004-6256/135/5/1877}, \href
  {https://ui.adsabs.harvard.edu/abs/2008AJ....135.1877E} {135, 1877}

\bibitem[\protect\citeauthoryear{{Freeman}}{{Freeman}}{1970}]{1970ApJ...160..811F}
{Freeman} K.~C.,  1970, \mn@doi [\apj] {10.1086/150474}, \href
  {https://ui.adsabs.harvard.edu/abs/1970ApJ...160..811F} {160, 811}

\bibitem[\protect\citeauthoryear{{Governato}, {Willman}, {Mayer}, {Brooks},
  {Stinson}, {Valenzuela}, {Wadsley}  \& {Quinn}}{{Governato}
  et~al.}{2007}]{2007MNRAS.374.1479G}
{Governato} F.,  {Willman} B.,  {Mayer} L.,  {Brooks} A.,  {Stinson} G.,
  {Valenzuela} O.,  {Wadsley} J.,   {Quinn} T.,  2007, \mn@doi [\mnras]
  {10.1111/j.1365-2966.2006.11266.x}, \href
  {https://ui.adsabs.harvard.edu/abs/2007MNRAS.374.1479G} {374, 1479}

\bibitem[\protect\citeauthoryear{{Hammer}, {Flores}, {Elbaz}, {Zheng}, {Liang}
  \& {Cesarsky}}{{Hammer} et~al.}{2005}]{2005A&A...430..115H}
{Hammer} F.,  {Flores} H.,  {Elbaz} D.,  {Zheng} X.~Z.,  {Liang} Y.~C.,
  {Cesarsky} C.,  2005, \mn@doi [\aap] {10.1051/0004-6361:20041471}, \href
  {https://ui.adsabs.harvard.edu/abs/2005A&A...430..115H} {430, 115}

\bibitem[\protect\citeauthoryear{{Hammer}, {Flores}, {Yang}, {Athanassoula},
  {Puech}, {Rodrigues}  \& {Peirani}}{{Hammer}
  et~al.}{2009}]{2009A&A...496..381H}
{Hammer} F.,  {Flores} H.,  {Yang} Y.~B.,  {Athanassoula} E.,  {Puech} M.,
  {Rodrigues} M.,   {Peirani} S.,  2009, \mn@doi [\aap]
  {10.1051/0004-6361:200810488}, \href
  {https://ui.adsabs.harvard.edu/abs/2009A&A...496..381H} {496, 381}

\bibitem[\protect\citeauthoryear{{Hopkins}, {Hernquist}, {Cox}, {Younger}  \&
  {Besla}}{{Hopkins} et~al.}{2008}]{2008ApJ...688..757H}
{Hopkins} P.~F.,  {Hernquist} L.,  {Cox} T.~J.,  {Younger} J.~D.,   {Besla} G.,
   2008, \mn@doi [\apj] {10.1086/592087}, \href
  {https://ui.adsabs.harvard.edu/abs/2008ApJ...688..757H} {688, 757}

\bibitem[\protect\citeauthoryear{{Hopkins}, {Cox}, {Younger}  \&
  {Hernquist}}{{Hopkins} et~al.}{2009}]{2009ApJ...691.1168H}
{Hopkins} P.~F.,  {Cox} T.~J.,  {Younger} J.~D.,   {Hernquist} L.,  2009,
  \mn@doi [\apj] {10.1088/0004-637X/691/2/1168}, \href
  {https://ui.adsabs.harvard.edu/abs/2009ApJ...691.1168H} {691, 1168}

\bibitem[\protect\citeauthoryear{{Larson} \& {Tinsley}}{{Larson} \&
  {Tinsley}}{1978}]{1978ApJ...219...46L}
{Larson} R.~B.,  {Tinsley} B.~M.,  1978, \mn@doi [\apj] {10.1086/155753}, \href
  {https://ui.adsabs.harvard.edu/abs/1978ApJ...219...46L} {219, 46}

\bibitem[\protect\citeauthoryear{{Miller} \& {Bregman}}{{Miller} \&
  {Bregman}}{2015}]{2015ApJ...800...14M}
{Miller} M.~J.,  {Bregman} J.~N.,  2015, \mn@doi [\apj]
  {10.1088/0004-637X/800/1/14}, \href
  {https://ui.adsabs.harvard.edu/abs/2015ApJ...800...14M} {800, 14}

\bibitem[\protect\citeauthoryear{{Naab} \& {Trujillo}}{{Naab} \&
  {Trujillo}}{2006}]{2006MNRAS.369..625N}
{Naab} T.,  {Trujillo} I.,  2006, \mn@doi [\mnras]
  {10.1111/j.1365-2966.2006.10252.x}, \href
  {https://ui.adsabs.harvard.edu/abs/2006MNRAS.369..625N} {369, 625}

\bibitem[\protect\citeauthoryear{{Naab} et~al.,}{{Naab}
  et~al.}{2014}]{2014MNRAS.444.3357N}
{Naab} T.,  et~al., 2014, \mn@doi [\mnras] {10.1093/mnras/stt1919}, \href
  {https://ui.adsabs.harvard.edu/abs/2014MNRAS.444.3357N} {444, 3357}

\bibitem[\protect\citeauthoryear{{Navarro}, {Frenk}  \& {White}}{{Navarro}
  et~al.}{1997}]{1997ApJ...490..493N}
{Navarro} J.~F.,  {Frenk} C.~S.,   {White} S. D.~M.,  1997, \mn@doi [\apj]
  {10.1086/304888}, \href
  {https://ui.adsabs.harvard.edu/abs/1997ApJ...490..493N} {490, 493}

\bibitem[\protect\citeauthoryear{{Oogi} \& {Habe}}{{Oogi} \&
  {Habe}}{2013}]{2013MNRAS.428..641O}
{Oogi} T.,  {Habe} A.,  2013, \mn@doi [\mnras] {10.1093/mnras/sts047}, \href
  {https://ui.adsabs.harvard.edu/abs/2013MNRAS.428..641O} {428, 641}

\bibitem[\protect\citeauthoryear{{Peschken} \& {{\L}okas}}{{Peschken} \&
  {{\L}okas}}{2019}]{2019MNRAS.483.2721P}
{Peschken} N.,  {{\L}okas} E.~L.,  2019, \mn@doi [\mnras]
  {10.1093/mnras/sty3277}, \href
  {https://ui.adsabs.harvard.edu/abs/2019MNRAS.483.2721P} {483, 2721}

\bibitem[\protect\citeauthoryear{{Peschken}, {Athanassoula}  \&
  {Rodionov}}{{Peschken} et~al.}{2017}]{2017MNRAS.468..994P}
{Peschken} N.,  {Athanassoula} E.,   {Rodionov} S.~A.,  2017, \mn@doi [\mnras]
  {10.1093/mnras/stx481}, \href
  {https://ui.adsabs.harvard.edu/abs/2017MNRAS.468..994P} {468, 994}

\bibitem[\protect\citeauthoryear{{Robertson}, {Bullock}, {Cox}, {Di Matteo},
  {Hernquist}, {Springel}  \& {Yoshida}}{{Robertson}
  et~al.}{2006}]{2006ApJ...645..986R}
{Robertson} B.,  {Bullock} J.~S.,  {Cox} T.~J.,  {Di Matteo} T.,  {Hernquist}
  L.,  {Springel} V.,   {Yoshida} N.,  2006, \mn@doi [\apj] {10.1086/504412},
  \href {https://ui.adsabs.harvard.edu/abs/2006ApJ...645..986R} {645, 986}

\bibitem[\protect\citeauthoryear{{Rodionov}, {Athanassoula}  \&
  {Peschken}}{{Rodionov} et~al.}{2017}]{2017A&A...600A..25R}
{Rodionov} S.~A.,  {Athanassoula} E.,   {Peschken} N.,  2017, \mn@doi [\aap]
  {10.1051/0004-6361/201628319}, \href
  {https://ui.adsabs.harvard.edu/abs/2017A&A...600A..25R} {600, A25}

\bibitem[\protect\citeauthoryear{{Rodriguez-Gomez} et~al.,}{{Rodriguez-Gomez}
  et~al.}{2017}]{2017MNRAS.467.3083R}
{Rodriguez-Gomez} V.,  et~al., 2017, \mn@doi [\mnras] {10.1093/mnras/stx305},
  \href {https://ui.adsabs.harvard.edu/abs/2017MNRAS.467.3083R} {467, 3083}

\bibitem[\protect\citeauthoryear{{Sales}, {Navarro}, {Theuns}, {Schaye},
  {White}, {Frenk}, {Crain}  \& {Dalla Vecchia}}{{Sales}
  et~al.}{2012}]{2012MNRAS.423.1544S}
{Sales} L.~V.,  {Navarro} J.~F.,  {Theuns} T.,  {Schaye} J.,  {White} S. D.~M.,
   {Frenk} C.~S.,  {Crain} R.~A.,   {Dalla Vecchia} C.,  2012, \mn@doi [\mnras]
  {10.1111/j.1365-2966.2012.20975.x}, \href
  {https://ui.adsabs.harvard.edu/abs/2012MNRAS.423.1544S} {423, 1544}

\bibitem[\protect\citeauthoryear{{Sauvaget}, {Hammer}, {Puech}, {Yang},
  {Flores}  \& {Rodrigues}}{{Sauvaget} et~al.}{2018}]{2018MNRAS.473.2521S}
{Sauvaget} T.,  {Hammer} F.,  {Puech} M.,  {Yang} Y.~B.,  {Flores} H.,
  {Rodrigues} M.,  2018, \mn@doi [\mnras] {10.1093/mnras/stx2453}, \href
  {https://ui.adsabs.harvard.edu/abs/2018MNRAS.473.2521S} {473, 2521}

\bibitem[\protect\citeauthoryear{{Snyder} et~al.,}{{Snyder}
  et~al.}{2015}]{2015MNRAS.454.1886S}
{Snyder} G.~F.,  et~al., 2015, \mn@doi [\mnras] {10.1093/mnras/stv2078}, \href
  {http://adsabs.harvard.edu/abs/2015MNRAS.454.1886S} {454, 1886}

\bibitem[\protect\citeauthoryear{{Sparre} \& {Springel}}{{Sparre} \&
  {Springel}}{2016}]{2016MNRAS.462.2418S}
{Sparre} M.,  {Springel} V.,  2016, \mn@doi [\mnras] {10.1093/mnras/stw1793},
  \href {https://ui.adsabs.harvard.edu/abs/2016MNRAS.462.2418S} {462, 2418}

\bibitem[\protect\citeauthoryear{{Sparre} \& {Springel}}{{Sparre} \&
  {Springel}}{2017}]{2017MNRAS.470.3946S}
{Sparre} M.,  {Springel} V.,  2017, \mn@doi [\mnras] {10.1093/mnras/stx1516},
  \href {https://ui.adsabs.harvard.edu/abs/2017MNRAS.470.3946S} {470, 3946}

\bibitem[\protect\citeauthoryear{{Springel}}{{Springel}}{2005}]{2005MNRAS.364.1105S}
{Springel} V.,  2005, \mn@doi [\mnras] {10.1111/j.1365-2966.2005.09655.x},
  \href {https://ui.adsabs.harvard.edu/abs/2005MNRAS.364.1105S} {364, 1105}

\bibitem[\protect\citeauthoryear{{Springel}}{{Springel}}{2010}]{2010MNRAS.401..791S}
{Springel} V.,  2010, \mn@doi [\mnras] {10.1111/j.1365-2966.2009.15715.x},
  \href {http://adsabs.harvard.edu/abs/2010MNRAS.401..791S} {401, 791}

\bibitem[\protect\citeauthoryear{{Springel} \& {Hernquist}}{{Springel} \&
  {Hernquist}}{2002}]{2002MNRAS.333..649S}
{Springel} V.,  {Hernquist} L.,  2002, \mn@doi [\mnras]
  {10.1046/j.1365-8711.2002.05445.x}, \href
  {https://ui.adsabs.harvard.edu/abs/2002MNRAS.333..649S} {333, 649}

\bibitem[\protect\citeauthoryear{{Springel} \& {Hernquist}}{{Springel} \&
  {Hernquist}}{2005}]{2005ApJ...622L...9S}
{Springel} V.,  {Hernquist} L.,  2005, \mn@doi [\apj] {10.1086/429486}, \href
  {https://ui.adsabs.harvard.edu/abs/2005ApJ...622L...9S} {622, L9}

\bibitem[\protect\citeauthoryear{{Tapia}, {Eliche-Moral}, {Aceves},
  {Rodr{\'\i}guez-P{\'e}rez}, {Borlaff}  \& {Querejeta}}{{Tapia}
  et~al.}{2017}]{2017A&A...604A.105T}
{Tapia} T.,  {Eliche-Moral} M.~C.,  {Aceves} H.,  {Rodr{\'\i}guez-P{\'e}rez}
  C.,  {Borlaff} A.,   {Querejeta} M.,  2017, \mn@doi [\aap]
  {10.1051/0004-6361/201628821}, \href
  {https://ui.adsabs.harvard.edu/abs/2017A&A...604A.105T} {604, A105}

\bibitem[\protect\citeauthoryear{{Taranu}, {Dubinski}  \& {Yee}}{{Taranu}
  et~al.}{2013}]{2013ApJ...778...61T}
{Taranu} D.~S.,  {Dubinski} J.,   {Yee} H.~K.~C.,  2013, \mn@doi [\apj]
  {10.1088/0004-637X/778/1/61}, \href
  {https://ui.adsabs.harvard.edu/abs/2013ApJ...778...61T} {778, 61}

\bibitem[\protect\citeauthoryear{{Toomre}}{{Toomre}}{1977}]{1977egsp.conf..401T}
{Toomre} A.,  1977, in {Tinsley} B.~M.,  {Larson} Richard B.~Gehret D.~C.,
  eds, Evolution of Galaxies and Stellar Populations. p.~401

\bibitem[\protect\citeauthoryear{{Toomre} \& {Toomre}}{{Toomre} \&
  {Toomre}}{1972}]{1972ApJ...178..623T}
{Toomre} A.,  {Toomre} J.,  1972, \mn@doi [\apj] {10.1086/151823}, \href
  {https://ui.adsabs.harvard.edu/abs/1972ApJ...178..623T} {178, 623}

\bibitem[\protect\citeauthoryear{{Vogelsberger} et~al.,}{{Vogelsberger}
  et~al.}{2014}]{2014MNRAS.444.1518V}
{Vogelsberger} M.,  et~al., 2014, \mn@doi [\mnras] {10.1093/mnras/stu1536},
  \href {http://adsabs.harvard.edu/abs/2014MNRAS.444.1518V} {444, 1518}

\bibitem[\protect\citeauthoryear{{de Vaucouleurs}}{{de
  Vaucouleurs}}{1948}]{1948AnAp...11..247D}
{de Vaucouleurs} G.,  1948, Annales d'Astrophysique, \href
  {https://ui.adsabs.harvard.edu/abs/1948AnAp...11..247D} {11, 247}

\makeatother
\end{thebibliography}

\section*{Appendix}

In this appendix we present a table describing the properties of all the merger cases for both samples used in this paper, namely the disc and elliptical samples. For each case, the table gives the ID of the remnant galaxy at redshift $z=0$, as well as its total, stellar and gaseous mass, the rotation to velocity dispersion values together with the S\'ersic indexes of the surface density fit (see section \ref{disc_sample_merg} and Fig. \ref{histg}), the redshift at which the major merger occured, the total, stellar and gaseous masses of both the progenitors before the merger (as in section \ref{disc_sample_merg}, we take the average over 3 snapshots where the distance is lower than 80 kpc), as well as their flatness. All the masses here are the ones provided by Illustris, but the flatness is the one we computed ourselves using the mass tensor, as in section \ref{res_disc}. Note that some cases have a total mass at $z=0$ lower than the sum of both progenitor masses before the merger. This seems surprising, so we checked those cases visually. We found that while right after the merger the remnant galaxy has a total mass of roughly the sum of both progenitor masses, some event later on strips part of the galaxy mass away. This is often the passage of a much more massive galaxy nearby that takes away mostly dark matter from the remnant galaxy.
\begin{table*}
  \caption{List and properties of the galaxies used in the paper: ID of the galaxy (column 1), sample it belongs to (column 2) total, stellar and gaseous mass (columns 3,4,5) as well as rotation to velocity dispersion and S\'ersic index values (columns 6,7) at redshift $z=0$, redshift at which the major merger occurs (column 8), followed by total, stellar and gaseous mass of both progenitors before the merger (columns 9,10,11,12,13,14), and flatness of both progenitors before the merger (columns 15,16). Masses are in $10^{10}M_{\odot}$.}
  
  \begin{sideways}
  \begin{tabular}{|c|c|c|c|c|c|c|c|c|c|c|c|c|c|c|c|}
    \hline
    1 & 2 & 3 & 4 & 5 & 6 & 7 & 8 & 9 & 10 & 11 & 12 & 13 & 14 & 15 & 16 \\
    ID  &  Sample  &  $M_{tot}$  &  $M_{*}$  &  $M_{gas}$  & $<v_{rot}>/\sigma_v$ & $n$  &  $z_{merg}$ &  $M_{tot,1}$ &   $M_{*,1}$ & $M_{gas,1}$  &  $M_{tot,2}$  & $M_{*,2}$  &  $M_{gas,2}$ & $flat_1$ & $flat_2$  \\
    &   &  ($z=0$)  &  ($z=0$)  &  ($z=0$)  & ($z=0$)&  ($z=0$) &   & &  & &    &  &  &  & \\
    \hline
    16948 & Disc & 99.825 & 11.211 & 7.428 & 0.536 & 0.702 & 1.07 & 74.873 & 1.283 & 12.219 & 20.602 & 1.087 & 4.361 & 0.742 & 0.761 \\ 
51817 & Disc & 164.83 & 19.81 & 0.217 & 0.416 & 1.62 & 1.3 & 166.127 & 5.509 & 19.995 & 72.71 & 5.062 & 10.377 & 0.496 & 0.55 \\ 
73684 & Disc & 25.037 & 7.722 & 0.902 & 0.65 & 0.821 & 0.7 & 59.687 & 4.186 & 7.884 & 19.12 & 0.685 & 3.637 & 0.589 & 0.718 \\ 
80741 & Disc & 69.922 & 6.903 & 0.188 & 0.475 & 1.127 & 1.36 & 49.513 & 1.589 & 6.657 & 57.24 & 1.854 & 6.505 & 0.826 & 0.575 \\ 
99155 & Disc & 46.06 & 5.617 & 4.677 & 0.711 & 1.143 & 1.15 & 31.707 & 0.795 & 7.036 & 8.316 & 0.324 & 1.697 & 0.809 & 0.819 \\ 
135292 & Disc & 78.389 & 5.675 & 0.176 & 0.507 & 1.372 & 0.89 & 43.813 & 2.001 & 3.588 & 99.907 & 1.847 & 8.805 & 0.864 & 0.89 \\ 
200655 & Disc & 76.694 & 12.723 & 1.801 & 0.5 & 1.746 & 1.15 & 93.933 & 3.43 & 11.839 & 172.951 & 4.289 & 21.115 & 0.719 & 0.579 \\ 
248239 & Disc & 190.058 & 11.481 & 0.883 & 0.542 & 1.318 & 1.3 & 113.957 & 3.756 & 12.396 & 28.165 & 0.546 & 5.318 & 0.794 & 0.69 \\ 
276191 & Disc & 13.028 & 5.227 & 0.441 & 0.428 & 1.008 & 1.11 & 28.14 & 1.316 & 3.856 & 105.416 & 1.359 & 10.857 & 0.854 & 0.785 \\ 
341073 & Disc & 322.412 & 8.041 & 6.917 & 0.545 & 1.443 & 0.38 & 72.894 & 3.39 & 4.909 & 102.403 & 2.585 & 5.814 & 0.803 & 0.67 \\ 
344090 & Disc & 401.151 & 16.209 & 1.47 & 0.577 & 2.064 & 0.35 & 138.707 & 8.361 & 7.14 & 244.312 & 6.225 & 15.569 & 0.684 & 0.601 \\ 
344195 & Disc & 302.99 & 18.628 & 6.76 & 0.692 & 1.483 & 0.2 & 102.453 & 7.705 & 8.622 & 163.412 & 8.37 & 13.487 & 0.862 & 0.726 \\ 
349189 & Disc & 280.983 & 10.869 & 6.126 & 0.755 & 1.13 & 1.3 & 79.111 & 1.968 & 11.801 & 21.141 & 0.661 & 4.255 & 0.764 & 0.782 \\ 
356914 & Disc & 276.044 & 10.115 & 9.189 & 0.486 & 0.836 & 0.6 & 118.199 & 1.885 & 11.228 & 32.805 & 1.729 & 5.386 & 0.644 & 0.556 \\ 
359796 & Disc & 198.726 & 12.398 & 1.94 & 0.735 & 0.8 & 1.53 & 66.031 & 1.139 & 11.232 & 20.378 & 0.789 & 3.93 & 0.75 & 0.645 \\ 
360893 & Disc & 267.217 & 13.807 & 11.878 & 0.617 & 0.511 & 1.0 & 93.69 & 1.914 & 10.688 & 23.501 & 0.892 & 4.231 & 0.719 & 0.815 \\ 
362307 & Disc & 291.227 & 6.126 & 5.716 & 0.504 & 1.217 & 1.47 & 69.582 & 1.588 & 7.633 & 18.083 & 0.596 & 3.505 & 0.844 & 0.856 \\ 
366317 & Disc & 299.428 & 11.148 & 4.223 & 0.478 & 1.705 & 0.38 & 110.657 & 4.847 & 8.069 & 93.371 & 1.782 & 5.854 & 0.822 & 0.73 \\ 
367696 & Disc & 161.32 & 8.138 & 10.781 & 0.376 & 1.601 & 0.82 & 12.95 & 0.6 & 3.262 & 33.199 & 0.244 & 5.74 & 0.801 & 0.656 \\ 
367785 & Disc & 272.747 & 15.533 & 0.34 & 0.595 & 1.142 & 1.41 & 108.889 & 1.547 & 14.91 & 33.448 & 1.653 & 6.327 & 0.886 & 0.807 \\ 
374189 & Disc & 190.82 & 6.154 & 4.025 & 0.37 & 1.01 & 1.21 & 78.998 & 1.301 & 7.999 & 24.19 & 1.313 & 2.803 & 0.814 & 0.76 \\ 
386479 & Disc & 214.462 & 7.671 & 9.253 & 0.936 & 1.973 & 0.5 & 121.231 & 3.309 & 9.556 & 54.115 & 1.786 & 5.115 & 0.746 & 0.756 \\ 
390096 & Disc & 186.778 & 5.945 & 7.723 & 0.483 & 1.06 & 0.33 & 50.522 & 1.69 & 4.593 & 74.135 & 1.895 & 7.393 & 0.619 & 0.779 \\ 
396467 & Disc & 167.031 & 6.952 & 7.271 & 0.439 & 0.962 & 0.5 & 88.181 & 2.854 & 12.709 & 26.018 & 1.038 & 4.368 & 0.678 & 0.741 \\ 
397431 & Disc & 140.273 & 5.413 & 9.724 & 0.819 & 0.504 & 0.95 & 45.441 & 0.549 & 6.347 & 11.876 & 0.32 & 2.635 & 0.806 & 0.757 \\ 
404068 & Disc & 121.094 & 5.844 & 11.119 & 0.712 & 0.531 & 0.68 & 57.943 & 1.287 & 7.773 & 20.868 & 0.771 & 4.824 & 0.848 & 0.694 \\ 
407728 & Disc & 121.833 & 5.669 & 3.36 & 0.528 & 1.197 & 1.3 & 57.697 & 1.328 & 7.306 & 25.04 & 0.763 & 5.548 & 0.813 & 0.708 \\ 
407890 & Disc & 150.595 & 5.857 & 6.349 & 0.881 & 1.36 & 0.7 & 64.91 & 2.018 & 7.661 & 41.315 & 1.669 & 4.992 & 0.742 & 0.594 \\ 
415149 & Disc & 122.597 & 6.64 & 5.485 & 0.492 & 1.019 & 0.46 & 68.507 & 1.695 & 8.838 & 33.822 & 1.292 & 6.274 & 0.658 & 0.747 \\ 
417612 & Disc & 100.99 & 5.239 & 8.867 & 0.688 & 0.597 & 1.15 & 27.796 & 0.439 & 5.51 & 23.589 & 0.427 & 3.933 & 0.687 & 0.883 \\ 
418069 & Disc & 128.818 & 7.129 & 1.104 & 0.426 & 1.315 & 0.95 & 60.289 & 2.15 & 7.614 & 33.958 & 1.238 & 7.048 & 0.763 & 0.76 \\ 
422233 & Disc & 122.024 & 5.409 & 4.049 & 0.32 & 0.968 & 0.48 & 49.451 & 1.923 & 3.91 & 57.632 & 1.311 & 7.235 & 0.877 & 0.749 \\ 
422540 & Disc & 105.234 & 6.945 & 5.823 & 0.338 & 0.722 & 0.46 & 62.891 & 2.212 & 7.989 & 22.346 & 1.298 & 4.979 & 0.88 & 0.676 \\ 
425211 & Disc & 120.203 & 5.114 & 3.274 & 0.323 & 0.752 & 1.25 & 21.333 & 1.097 & 4.453 & 52.69 & 1.296 & 7.43 & 0.823 & 0.801 \\ 
425559 & Disc & 107.632 & 5.558 & 5.602 & 0.432 & 0.969 & 0.38 & 22.847 & 1.925 & 3.649 & 75.413 & 1.883 & 9.171 & 0.683 & 0.724 \\ 
427698 & Disc & 93.71 & 5.707 & 7.905 & 0.566 & 0.539 & 0.82 & 46.788 & 0.605 & 6.96 & 12.625 & 0.452 & 2.976 & 0.675 & 0.716 \\ 
430982 & Disc & 101.041 & 5.141 & 3.554 & 0.436 & 0.938 & 1.15 & 51.349 & 1.618 & 6.497 & 16.837 & 0.731 & 3.316 & 0.756 & 0.857 \\ 
436092 & Disc & 80.677 & 5.919 & 6.083 & 0.583 & 0.434 & 0.95 & 43.796 & 1.102 & 7.011 & 12.632 & 0.457 & 3.137 & 0.661 & 0.725 \\
11 & Elliptical & 99.184 & 31.903 & 0.01 & 0.014 & 1.825 & 0.52 & 202.815 & 20.535 & 3.187 & 170.609 & 20.411 & 4.276 & 0.543 & 0.702 \\ 
16960 & Elliptical & 54.877 & 5.747 & 0.0 & 0.082 & 1.887 & 0.79 & 140.152 & 3.56 & 8.321 & 69.034 & 1.637 & 4.932 & 0.898 & 0.916 \\ 
59397 & Elliptical & 26.172 & 6.212 & 0.001 & 0.029 & 1.555 & 0.48 & 74.699 & 4.839 & 4.341 & 24.511 & 2.829 & 4.199 & 0.936 & 0.713 \\ 
80735 & Elliptical & 2006.591 & 55.283 & 1.425 & 0.08 & 3.247 & 0.27 & 754.693 & 22.01 & 1.934 & 452.345 & 23.537 & 1.212 & 0.636 & 0.626 \\ 
104805 & Elliptical & 128.92 & 11.701 & 0.009 & 0.014 & 1.764 & 0.4 & 461.376 & 9.079 & 3.04 & 261.698 & 4.134 & 0.761 & 0.919 & 0.711 \\ 
204398 & Elliptical & 58.34 & 11.256 & 0.122 & 0.151 & 1.46 & 0.46 & 67.997 & 5.159 & 5.918 & 143.91 & 7.414 & 8.169 & 0.802 & 0.689 \\ 
225517 & Elliptical & 1820.185 & 31.482 & 0.78 & 0.001 & 2.359 & 0.89 & 781.146 & 14.279 & 18.098 & 453.146 & 15.841 & 8.174 & 0.658 & 0.671 \\ 
    \hline
  \end{tabular}
  \end{sideways}
  \label{table1}
  \end{table*}

\end{document}